\documentclass[reqno,11pt]{article}
\pdfoutput=1
\usepackage{jheppub}
\usepackage{extarrows}
\usepackage{epsfig}
\usepackage{amssymb}
\usepackage{verbatim}
\usepackage{amsmath}
\usepackage{stmaryrd}
\usepackage{mathrsfs}
\usepackage{hyperref}
\usepackage{dsfont}
\usepackage{slashed}
\usepackage[dvipsnames]{xcolor}
\usepackage{epstopdf}
\usepackage{eucal}
\usepackage{graphicx}
\usepackage{color}
\usepackage{bbold}
\usepackage[version=3]{mhchem}
\usepackage{shuffle}
\usepackage[all,cmtip,2cell]{xy}

\newcommand*{\sbullet}{\raisebox{0.1ex}{\scalebox{0.6}{$\bullet$}}}

\def\ZZ{{\mathbb{Z}}}

\def\RR{{\mathbb{R}}}

\def\gfrak{{\mathfrak{g}}}

\DeclareMathAlphabet{\mathsfit}{T1}{\sfdefault}{\mddefault}{\sldefault}
\SetMathAlphabet{\mathsfit}{bold}{T1}{\sfdefault}{\bfdefault}{\sldefault}

\def\LBA{{\mathfrak{L}_{\mathrm{BA}}}}
\def\LYM{{\mathfrak{L}_{\mathrm{YM}}}}

\def\ud{{\mathrm{d}}}
\def\ui{{\mathrm{i}}}
\def\id{{\mathrm{id}}}

\def\to{\rightarrow}
\def\lto{\longrightarrow}

\def\GF{{G^{\mathrm{F}}}}

\newcommand\tr{\operatorname{tr}}
\newcommand\ue{\operatorname{e}}
\newcommand\im{\operatorname{im}}
\newcommand\uud{\operatorname{d}}

\newcommand{\ba}{\begin{eqnarray}}
\newcommand{\ea}{\end{eqnarray}}

\DeclareMathAlphabet{\mathbfit}{OML}{cmm}{b}{it}

\title{$L_\infty$-algebras and the perturbiner expansion}

\author{Cristhiam Lopez-Arcos${}^{a}$ and Alexander Quintero V\'{e}lez${}^{b}$}

\affiliation{${}^{a}$Grupo de Electr\'onica y Automatizaci\'on, Instituci\'on universitaria Salazar y Herrera, \\ \phantom{${}^{a}$}Carrera 70 $\#$ 52--49, Medell\'in, Colombia\\
${}^{b}$Escuela de Matem\'{a}ticas, Universidad Nacional de Colombia Sede Medell\'{i}n, \\ \phantom{${}^{b}$}Carrera 65 $\#$ 59A--110, Medell\'{i}n, Colombia}

\emailAdd{crismalo@gmail.com, aquinte2@unal.edu.co}

\abstract{Certain classical field theories admit a formal multi-particle solution, known as the perturbiner expansion, that serves as a generating function for all the tree-level scattering amplitudes and the Berends-Giele recursion relations they satisfy. In this paper it is argued that the minimal model for the $L_{\infty}$-algebra that governs a classical field theory contains enough information to determine the perturbiner expansion associated to such theory.  This gives a prescription for computing the tree-level scattering amplitudes by inserting the perturbiner solution into the homotopy Maurer-Cartan action for the $L_{\infty}$-algebra. We confirm the method in the non-trivial examples of bi-adjoint scalar and Yang-Mills theories. 
}

\begin{document}

\maketitle

%%%%%%%%%%%%%%%%%%%%%%%%%%%%%%%%%%
\section{Introduction} 
$L_{\infty}$-algebras are natural generalisations of graded Lie algebras, in which the Jacobi identity is allowed to hold only up to homotopy. They were introduced by Schlessinger and Stasheff in \cite{Schlessinger-Stasheff1985} in the context of rational homotopy theory. For several years now, $L_{\infty}$-algebras have become increasingly important in mathematical physics, especially in string field theory \cite{Witten:1992yj,Zwiebach:1992ie}, where they organize the terms of higher order in perturbed actions, and in deformation quantisation \cite{Cattaneo:1999fm,Kontsevich:1997vb}.

The main focus of this article will be on $L_{\infty}$-algebras that appear naturally in the Batalin-Vilkovisky quantisation of classical field theories. This has been the subject matter of several recent works \cite{Hohm:2017pnh,Hohm:2017cey,Jurco:2019bvp,Jurco:2018sby}. The guiding philosophy is that the equations of motion of a classical field theory can be written in the form of homotopy Maurer-Cartan
equations associated with an appropriate $L_{\infty}$-algebra. These equations are ``universal'' in the sense that they can be derived from the variation of an action functional which is a higher order version of the Chern-Simons action. Somewhat more generally, the $L_{\infty}$-algebra underlying a classical field theory captures all the information about its gauge symmetries, field content and Noether currents.

Recently, it has been established by Macrelli, S\"{a}mann and Wolf \cite{Macrelli:2019afx} that the $L_{\infty}$-structure of a classical field theory may also be used to determine the recursion relations for its tree-level scattering amplitudes (see also \cite{Nutzi:2018vkl} and \cite{Arvanitakis:2019ald}). These authors worked out in detail the concrete case of Yang-Mills theory. More precisely, they found that the recursion relations for the tree-level scattering amplitudes in Yang-Mills theory, which are known as the Berends-Giele recursion relations \cite{Berends:1987me}, arise as recursion relations of the underlying quasi-isomorphism between the Yang-Mills $L_{\infty}$-algebra and its minimal model. 

On the other hand, it is well-known that, given a massless quantum field theory, a knowledge of the tree-level scattering amplitudes implies a knowledge of all solutions of the classical equations of motion of the corresponding classical field theory. To substantiate this statement, Rosly and Selivanov \cite{Rosly:1996vr,Rosly:1997ap,Selivanov:1997aq,Selivanov:1997ts,Rosly:1998vm,Selivanov:1998hn} considered an ansatz for such solutions as a formal plane-wave expansion, which, at the same time, can be regarded as a generating function for all tree-level scattering amplitudes in the theory. In accordance with these authors' terminology, this ansatz is called the pertrubiner expansion. 

In view of the preceding discussion, it should not come as a surprise that the perturbiner expansion for a classical field theory can be systematically encoded in the minimal model of the $L_{\infty}$-algebra that governs it. It is the purpose of this article to corroborate the validity of this statement in the concrete cases of bi-adjoint scalar and Yang-Mills theories. To be more specific, for these theories, we shall prove that the correct $L_{\infty}$-structure on the minimal models can be constructed explicitly, and this goes hand in hand with a derivation of the perturbiner expansions ``from first principles'', i.e., from the defining $L_{\infty}$-structures. As a by product, we shall also show that the actual tree-level scattering amplitudes can obtained by plugging in directly the perturbiner expansions into the homotopy Maurer-Cartan actions. This is to be contrasted with the original deduction of these perturbiner expansions \cite{Mafra:2016ltu,Mizera:2018jbh}, where they are merely thought of as ansatzs that lead to a series of recursion relations, such that finding a solution to the recurrence leads to finding a solution to the equations of motion. It could therefore appear that by stressing the abstract algebraic $L_{\infty}$-structure of the theories in question one gains a deeper conceptual understanding. Morever, this approach to determining perturbiner expansions seems to be universal and applicable to any classical field theory admitting an $L_{\infty}$-algebra formulation.

To close this introduction, we wish to emphasise that perturbiner methods have been the object of much recent attention. Mafra and Schlotterer have used them extensively to study perturbative aspects of $D=10$ super Yang-Mills theories \cite{Mafra:2015gia,Mafra:2015vca,Lee:2015upy}, as well as $\alpha'$-expansions of disk integrals in string theory \cite{Mafra:2016mcc}. They have also been applied to derive Berends-Giele recursion relations for the bi-adjoint scalar theory \cite{Mafra:2016ltu} and the study of $\alpha'$-deformations of Yang-Mills theory \cite{Garozzo:2018uzj}. In addition to all these, Mizera and Skrzypek have constructed perturbiner expansions for effective field theories with and without color, including non-linear sigma models, special Galileon theory, and Born-Infeld theory \cite{Mizera:2018jbh}. Lastly, and even more recently, the perturbiner expansion has been applied to obtain multiparticle super Yang-Mills superfields in the BCJ gauge \cite{Bridges:2019siz}. 

The present article is organized as follows. We begin in Section \ref{sec:2} with a review of the basic definitions and results concerning $L_{\infty}$-algebras, homotopy Maurer-Cartan theory, and the perturbiner expansions for the bi-adjoint scalar and Yang-Mills theories. In Section \ref{sec:3} we discuss the $L_{\infty}$-algebras relevant to the bi-adjoint scalar and Yang-Mills theories and demonstrate how the perturbiner expansions for such theories arise from the $L_{\infty}$-structure of the corresponding minimal models. Section \ref{sec:4} describes how the tree-level scattering amplitudes in bi-adjoint scalar and Yang-Mills theories can be obtained by inserting the perturbiner expansions into their respective homotopy Maurer-Cartan actions. The summary and conclusions are given in Section \ref{sec:5}.

%%%%%%%%%%%%%%%%%%%%%%%%%%%%%%%%%%
\section{Preliminaries}\label{sec:2}
As alluded to in the introduction, our interpretation of the perturbiner expansion for the bi-adjoint scalar and Yang-Mills theories is tied up in the $L_{\infty}$-algebra language.  We therefore begin with a review of the required facts. 

\subsection{$L_\infty$-algebras}\label{sec:2.1}
More information on this subject can be found in \cite{Stasheff1992,Lada-Markl1995,Kajiura:2003ax} and the references therein. The relevance of $L_{\infty}$-algebras in physics was discovered in \cite{Witten:1992yj} and is explained in detail in \cite{Zwiebach:1992ie}. 

Let $V = \bigoplus_{k \in \ZZ} V^{k}$ a $\ZZ$-graded vector space. Given a permutation $\sigma \in \mathfrak{S}_n$ and homogenoeus elements $v_1,\dots,v_n \in V$, we define the \emph{graded Koszul sign} $\chi(\sigma;v_1,\dots,v_n)$ to be the product of the signature of the permutation $(-1)^{\sigma}$ with a factor $(-1)^{\vert v_i \vert  \vert v_{i+1}\vert }$ for each transposition of $v_i$ and $v_{i+1}$ involved in the permutation. As a piece of terminology, we recall also that a permutation $\sigma \in \mathfrak{S}_n$ is called an $(i,n-i)$\emph{-shuffle} if it satisfies the inequalities $\sigma(1) < \cdots < \sigma(i)$ and $\sigma(i+1) < \cdots < \sigma(n)$. The set of $(i,n-i)$-unshuffles is denoted by $\mathfrak{S}_{i,n-i}$. More generally, a $(i_1,\dots,i_r)$-shuffle means a permutation $\sigma \in \mathfrak{S}_{n}$ with $n= i_1 + \cdots + i_r$ such that the order is preserved within each block of lenght $i_1,\dots,i_r$. The set consisting of all such shuffles we denote by $\mathfrak{G}_{i_1,\dots,i_r}$. 

An $L_{\infty}$\emph{-algebra} is a $\ZZ$-graded vector space $L = \bigoplus_{k \in \ZZ} L^{k}$ equipped with linear maps $l_n \colon L^{\otimes n} \to L$ of degree $2-n$ which are totally graded skew symmetric in the sense that 
\begin{equation}
l_n (x_{\sigma(1)},\dots,x_{\sigma(n)}) = \chi(\sigma; x_1,\dots,x_n) l_n (x_1,\dots,x_n),
\end{equation}
for any $\sigma \in \mathfrak{S}_n$ and $x_1,\dots,x_n \in L$, and are also required to satisfy the constraints
\begin{equation}\label{eqn:2.2}
\sum_{i=1}^{n}(-1)^{n-i} \sum_{\sigma \in \mathfrak{S}_{i,n-i}}  \chi(\sigma;x_1,\dots,x_n) l_{n-i+1} (l_{i}(x_{\sigma(1)},\dots,x_{\sigma(i)}), x_{\sigma(i+1)},\dots,x_{\sigma(n)}) = 0,
\end{equation}
for any $n \geq 1$ and $x_1,\dots,x_n \in L$.

Condition \eqref{eqn:2.2} may appear somewhat mysterious, but actually provides a generalisation of the Jacobi identity for an ordinary Lie algebra. Let us try to understand this by examining it in particular cases. For $n = 1$, it states that $l_1$ is of degree $1$ and satisfies
\begin{equation*}
l_1(l_1(x)) = 0,
\end{equation*}
for all $x \in L$. This implies that we have a cochain complex of vector spaces
$$
\cdots \xlongrightarrow{l_1} L^{k-1} \xlongrightarrow{l_1} L^{k} \xlongrightarrow{l_1} L^{k+1} \xlongrightarrow{l_1} \cdots.
$$
For $n = 2$, we have that $l_2$ is of degree $0$ and satisfies
\begin{equation*}
l_1(l_2(x_1,x_2)) = l_2(l_1(x_1),x_2) + (-1)^{\vert x_1\vert\vert x_2 \vert} l_2(l_1(x_2),x_1),
\end{equation*}
for all $x_1,x_2 \in L$. Thus $l_2$ induces a binary operation on $L$ and $l_1$ is a derivation with respect to $l_2$. Finally, $n=3$ yields
\begin{align*}
\begin{split}
 l_2(l_2(x_1,x_2),x_3) &+ (-1)^{\vert x_1\vert ( \vert x_2\vert +\vert x_3\vert)} l_2(l_2(x_2,x_3),x_1)  + (-1)^{\vert x_2 \vert ( \vert x_1\vert +\vert x_3\vert)}   l_2(l_2(x_1,x_3),x_2) \\
 &= -l_1 (l_3 (x_1,x_2,x_3)),
\end{split}
\end{align*}
for all $x_1,x_2,x_3 \in L$ with $l_1 (x_1) = l_1(x_2) = l_1(x_3) = 0$. This means that $l_2$ satisfies the graded Jacobi identity up to homotopy, but more is true:~the homotopy is provided by $l_3$, which is built into the definition of $L$. 

One should note that grading is essential to nontrivial $L_{\infty}$-algebras. An $L_{\infty}$-algebra concentrated in degree $0$ is necessarily a Lie algebra (all $l_n$ vanish for $n \neq 2$). 

For a pair of $L_{\infty}$-algebras $L$ and $L'$ there is  a natural notion of $L_{\infty}$-morphism from $L$ to $L'$. Namely, such a morphism consists of the data $f=(f_n)_{n \geq 1}$ where $f_n \colon L^{\otimes n} \to L'$ is a linear totally graded skew symmetric map of degree $1-n$ such that
\begin{align}\label{eqn:2.3}
\begin{split}
&\sum_{i=1}^{n} (-1)^{n-i} \sum_{\sigma \in \mathfrak{S}_{i, n-i}} \chi(\sigma; x_1,\dots,x_n) f_{n-i+1}(l_i (x_{\sigma(1)},\dots, x_{\sigma(i)} ), x_{\sigma(i+1)},\dots, x_{\sigma(n)}) \\
&= \sum_{r = 1}^{n}\frac{1}{r!} \sum_{i_1+\dots+i_r = n} \sum_{\sigma \in \mathfrak{S}_{i_1,\dots,i_r}} \chi(\sigma; x_1,\dots,x_n) \zeta(\sigma;  x_1,\dots,x_n) \\
&\qquad \qquad \qquad\qquad \qquad \times l'_r (f_{i_1}(x_{\sigma(1)},\dots, x_{\sigma(i_1)}), \dots, f_{i_r}(x_{\sigma(i_1+\cdots + i_{r-1} + 1)},\dots, x_{\sigma(n)})),
\end{split}
\end{align}
for any $n \geq 1$ and $x_1,\dots, x_n \in L$. The sign $\zeta(\sigma; x_1,\dots, x_n)$ on the right is given by
\begin{equation}
\zeta(\sigma; x_1,\dots, x_n) = (-1)^{\sum_{1\leq p < q \leq r} i_p i_q +  \sum_{q=1}^{r-1} i_q (r-q) + \sum_{p=2}^{r} (1-i_p) \sum_{q=1}^{i_1 + \dots + i_{p-1}}\vert x_{\sigma(q)}\vert} . 
\end{equation}
For $n = 1$ this yields the following condition
\begin{equation*}
f_1 (l_1(x)) = l'_1 (f_1(x)),
\end{equation*}
for all $x \in L$. For $n=2$ the condition reads
\begin{align*}
\begin{split}
l'_2(f_1(x_1),f_1(x_2)) &- f_1(l_2 (x_1,x_2))  \\
&= - l'_1(f_2(x_1,x_2)) + f_2(l_1(x_1),x_2) + (-1)^{(\vert x_1 \vert + 1)(\vert x_2\vert +1)}f_2 (l_1(x_2),x_1),
\end{split}
\end{align*}
for all $x_1,x_2 \in L$. The first equation implies that $f_1$ defines a morphism of complexes. The second equation implies that $f_1$ preserves the binary operation given by $l_2$ up to a homotopy given by $f_2$. More generally one might say that $f=(f_n)_{n \geq 1}$ preserves the $l_n$ up to homotopy. 

An $L_{\infty}$-morphism $f$ is called an $L_{\infty}$-\emph{quasi-isomorphism} if $f_1$ is a quasi-isomorphism. Two $L_{\infty}$-algebras $L$ and $L'$ are said to be $L_{\infty}$-quasi-isomorphic as $L_{\infty}$-algebras if there is an $L_{\infty}$-morphism $f \colon L \to L'$ that is an $L_{\infty}$-quasi-isomorphism. 

We now come to a result that allows us to pass to the cohomology of an $L_{\infty}$-algebra without losing too much information. Let $L$ be an $L_{\infty}$-algebra. As noted before, $l_1$ gives $L$ the structure of a cochain complex, and we may take cohomology to yield $H^{\sbullet}(L)$. By choosing representatives of each cohomology class we may define an embedding $i \colon H^{\sbullet}(L) \hookrightarrow L$. Thanks to a theorem of Kadeishvili~\cite{Kadeishvili1982}, we may define an $L_{\infty}$-structure on $H^{\sbullet}(L)$ such that $l'_1=0$, and there is a $L_{\infty}$-quasi-isomorphism $f$ from $H^{\sbullet}(L)$ to $L$ with $f_1$ equal to the embedding $i$. Here, $l'_1$ refers to the $L_{\infty}$-structure on $H^{\sbullet}(L)$. This $L_{\infty}$-structure is not unique, but it is unique up to $L_{\infty}$-isomorphisms. An $L_{\infty}$-algebra with $l_1=0$ is called a \emph{minimal} $L_{\infty}$-algebra; thus the above may be interpreted as saying that each $L_{\infty}$-algebra has an essentially unique minimal model. 

It is relatively easy to construct the minimal model in practice. A rather simple example of an $L_{\infty}$-algebra is given by $l_{n}=0$ for $n \geq 3$. Such an algebra is called a \emph{DG Lie algebra} (where DG stands for ``differential graded''). In this paper, we will need to put an $L_{\infty}$-structure on the cohomology of a DG Lie algebra, which may be done explicitly as follows. Suppose we define a projection $p \colon L \to H^{\sbullet}(L)$ such that $p \circ i = \mathrm{id}_{H^{\sbullet}(L)}$ and furthermore assume that we have a contracting homotopy $h \colon L \to L$. The latter means that $h$ is a map of degree $-1$ such that $\id_{L} - i \circ p = l_1 \circ h + h \circ l_1$. Then the $L_{\infty}$-quasi-isomorphism between $H^{\sbullet}(L)$ and $L$ is determined by the maps $f_n \colon H^{\sbullet}(L)^{\otimes n} \to L$ which are constructed recursively as
\begin{align}\label{eqn:2.5}
\begin{split}
f_1 (x_1) &= i(x_1), \\
f_2(x_1,x_2) &= -(h \circ l_2)(f_1(x_1),f_1(x_2)), \\
&\,\,\,\vdots \\
f_n (x_1,\dots,x_n) &=  -  \tfrac{1}{2} \sum_{i =1}^{n-1} \sum_{\sigma \in \mathfrak{S}_{i, n-i}} \chi(\sigma; x_1,\dots, x_n)  \\
& \qquad \qquad \qquad \quad \times (h \circ l_2) (f_i (x_{\sigma (1)} ,\dots, x_{\sigma(i)} ), f_{n-i}(x_{\sigma (i+1)} ,\dots, x_{\sigma(n)})), 
\end{split}
\end{align}
for all $x_1,\dots, x_n \in H^{\sbullet}(L)$. Likewise, the higher order brackets $l'_n \colon H^{\sbullet}(L)^{\otimes n} \to H^{\sbullet}(L)$ are given by
\begin{align}\label{eqn:2.6}
\begin{split}
l'_1 (x_1) &= 0, \\
l'_2(x_1,x_2) &=  (p \circ l_2)(f_1(x_1),f_1(x_2)), \\
&\,\,\,\vdots \\
l'_n (x_1,\dots,x_n) &= \tfrac{1}{2} \sum_{i =1}^{n-1} \sum_{\sigma \in \mathfrak{S}_{i, n-i}} \chi(\sigma; x_1,\dots, x_n)  \\
& \qquad \qquad \quad \quad\, \times (p \circ l_2) (f_i (x_{\sigma (1)} ,\dots, x_{\sigma(i)} ), f_{n-i}(x_{\sigma (i+1)} ,\dots, x_{\sigma(n)}))  ,
\end{split}
\end{align}
for all $x_1,\dots, x_n \in H^{\sbullet}(L)$. We refer to the Appendix A of \cite{Macrelli:2019afx} for more details. 

To complete our brief exposition on $L_{\infty}$-algebras, we need to introduce one more ingredient. By a \emph{cyclic inner product} on an $L_{\infty}$-algebra $L$ we mean a non-degenerate graded symmetric bilinear map $\langle , \rangle \colon L \times L \to \RR$  of degree $k$ such that
\begin{equation}\label{eq:2.7}
\langle x_1, l_{n}(x_2,\dots,x_{n+1}) \rangle = (-1)^{n + n(\vert x_1 \vert + \vert x_{n+1} \vert) + \vert x_{n+1}\vert \sum_{i=1}^{n} \vert x_i \vert}\langle x_{n+1}, l_{n}(x_1,\dots,x_{n}) \rangle,
\end{equation}
for any $n \geq 1$ and $x_1,\dots,x_{n+1} \in L$. An $L_{\infty}$-algebra equipped with a cylic inner product will be called a \emph{cyclic $L_{\infty}$-algebra}.

Given two cyclic $L_{\infty}$-algebras $L$ and $L'$, an $L_{\infty}$-morphism $f \colon L \to L'$ is itself called \emph{cyclic} if it satisfies the supplementary conditions
\begin{equation}
\langle f_1 (x_1), f_1 (x_2) \rangle' = \langle x_1, x_2 \rangle,
\end{equation}
and 
\begin{equation}
\sum_{i + j = n} \langle f_i (x_1,\dots, x_i), f_j (x_{i+1},\dots, x_{n}) \rangle' = 0,
\end{equation}
for any $n \geq 3$ and $x_1,\dots, x_{n} \in L$. With this definition, it can be shown that the minimal model theorem extends to cyclic $L_{\infty}$-algebras. For further details on this, see again the Appendix A of~\cite{Macrelli:2019afx}. 

%%%%%%%%%%%%%%%%%%%%%%%%%%%%%%%%

\subsection{Homotopy Maurer-Cartan theory}
To an $L_{\infty}$-algebra is associated a field theory known as the homotopy Maurer-Cartan theory.  This theory should be thought of as a broad generalisation of Chern-Simons theory. In what follows, we shall only sketch the details and refer to \cite{Jurco:2018sby} (see also \cite{Macrelli:2019afx}). 

Let $L$ be an $L_{\infty}$-algebra with higher order brackets $l_n$. An element $a \in L^{1}$ is said to be a \emph{Maurer-Cartan element} if  
\begin{equation}\label{eq:2.8}
\sum_{n \geq 1} \frac{1}{n!} l_n (a,\dots,a) = 0. 
\end{equation}
This equation, which describes an abstract form of ``flatness'', is known as the \emph{homotopy Maurer-Cartan equation}. It will play a key role in our considerations below. 

In general, an $L_{\infty}$-algebra structure can be deformed by its Maurer-Cartan elements. In order to put this a bit more precisely, let $L$ be an $L_{\infty}$-algebra and let $a$ be a Maurer-Cartan element in it. Then one can consider a new sequence of brackets on $L$ given by the formula
\begin{equation}
l^{a}_n (x_1,\dots,x_n) = \sum_{k \geq 0} \frac{1}{k!} l_{n+k} (a,\dots,a,x_1,\dots,x_n),
\end{equation}
for all $x_1,\dots,x_n \in L$. It is known (see, e.g., \cite{Getzler2009}) that the underlying $\ZZ$-graded vector space $L$ equipped with the higher order brackets $l^{a}_n$ is again an $L_{\infty}$-algebra. In fact, the constraints imposed by \eqref{eqn:2.2} imply the homotopy Maurer-Cartan equation \eqref{eq:2.8}.  

We next consider how Maurer-Cartan elements behave under $L_{\infty}$-morphisms. To this end, let $L$ and $L'$ be two $L_{\infty}$-algebras and let $f \colon L \to L'$ be an $L_{\infty}$-morphism between them. Under such a morphism, an arbitrary element $a \in L^{1}$ transforms according to 
\begin{equation}
a \mapsto a' = \sum_{n \geq 1} \frac{1}{n!} f_{n}(a,\dots,a). 
\end{equation}
Furthermore, it is not hard to see that
\begin{equation}
\sum_{n \geq 1} \frac{1}{n!} l'_n (a',\dots, a') = \sum_{k \geq 0} \frac{1}{k!} f_{k+1} \left( a,\dots, a , \sum_{n \geq 1} \frac{1}{n!} l_n (a,\dots,a) \right). 
\end{equation}
As a consequence, if $a$ is a Maurer-Cartan element in $L$, then $a'$ is a Maurer-Cartan element in $L'$. Thus, Maurer-Cartan elements are preserved under $L_{\infty}$-morphisms. 

Now to the point. Let $L$ be a cyclic $L_{\infty}$-algebra with cyclic inner product of degree $-3$. Then it turns out that the homotopy Maurer-Cartan equation \eqref{eq:2.8} can be derived from a variational principle. The action functional that describes it is given by
\begin{equation}\label{eq:2.9}
S_{\mathrm{MC}}[a] = \sum_{n \geq 1} \frac{1}{(n+1)!} \langle a, l_{n}(a,\dots,a) \rangle. 
\end{equation}
Indeed, using the cyclicity property \eqref{eq:2.7}, it is straightforward to check that the critical points of this functional are the solutions to the homotopy Maurer-Cartan equation \eqref{eq:2.8}. We shall refer to \eqref{eq:2.9} as the \emph{homotopy Maurer-Cartan action}. 

It is noteworthy that the homotopy Maurer-Cartan action \eqref{eq:2.9} is invariant under a set of infinitesimal transformations of the form
\begin{equation}
\delta_{c_0} a = \sum_{n \geq 0} \frac{1}{n!} l_{n+1}(a,\dots,a,c_0). 
\end{equation}
with infinitesimal parameters $c_0 \in L^{0}$. The explicit form of this transformations is of independent interest, for example for a better understanding of the moduli space of Maurer-Cartan elements for $L$. For a thorough discussion, see \cite{Getzler2009}. 

%%%%%%%%%%%%%%%%%%%%%%%%%%%%%%%%

\subsection{Bi-adjoint scalar theory and its perturbiner expansion}\label{sec:2.3}
This subsection reviews some basic features of the bi-adjoint scalar theory and the perturbiner expansion for the solution of its non-linear field equations. For further details we refer the reader to \cite{BjerrumBohr:2012mg,Cachazo:2013iea,Mafra:2016ltu,Mizera:2018jbh}. 

Before beginning, some comments about our notation. We let $\RR^{1,d-1}$ be the $d$-dimensional Minkowski spacetime. We take standard coordinates $x^{0}, x^{1},\dots,x^{d-1}$, where $x^{0}$ represents time. The metric tensor $\eta_{\mu\nu}$ is diagonal, with elements $\eta_{00}=-1$, $\eta_{\mu\nu} = 1$ if $\mu,\nu = 1,\dots, d-1$ and $\eta_{\mu\nu} = 0$ if $\mu \neq \nu$, and the standard volume element is
\begin{equation}
\ud^{d} x = \ud x^{0} \wedge \ud x^{1} \wedge  \cdots \wedge  \ud x^{d-1}.
\end{equation}
Spacetime coordinate labels are lowered and raised by using $\eta_{\mu\nu}$ and its inverse $\eta^{\mu\nu}$, respectively. The usual convention that repeated indices are summed over is used throughout. If $v = (v^0, v^1,\dots, v^{d-1})$ and $w = (w^0, w^1,\dots, w^{d-1})$ are vectors in $\RR^{1,d-1}$, their inner product is denoted by 
\begin{equation}
v \cdot w =  \eta_{\mu\nu} v^{\mu}w^{\nu} = -v^0 w^0 + v^1 w^1 + \cdots + v^{d-1} w^{d-1}.
\end{equation}
The d'Alembertian operator is defined as
\begin{equation}
\square = \eta^{\mu\nu}\frac{\partial^2}{\partial x^{\mu}\partial x^{\nu}}= -\frac{\partial^2}{(\partial x^{0})^2} + \frac{\partial^2}{(\partial x^{1})^2} + \cdots + \frac{\partial^2}{(\partial x^{d-1})^2}.
\end{equation}
We shall also use the shorthand notation $\partial_{\mu}$ for the partial derivative $\partial/ \partial x^{\mu}$.

The bi-adjoint scalar theory is a $d$-dimensional massless theory with a cubic interaction whose fundamental field transforms bi-linearly in the adjoint representation of two independent global symmetries. More precisely, the theory is specified by the choice of two compact semi-simple Lie groups $G$ and $G'$ whose corresponding Lie algebras will be denoted by $\gfrak$ and $\gfrak'$. We pick generators $T^a$ and $T^{\prime a'}$ for $\gfrak$ and $\gfrak'$ respectively, and, following the customary convention, let the associated structure constants be given by $[T^a,T^b] = \ui f^{ab}_{\phantom{ab}c}T^c$ and $[T^{\prime a'},T^{\prime b'}] = \ui f^{\prime a'b'}_{\phantom{a'b'}c'}T^{\prime c'}$. We also let $\kappa^{ab} = \tr (T^{a}T^{b})$ and $\kappa'^{a'b'} = \tr(T'^{a'}T'^{b'})$ be the components of the Cartan-Killing forms on $\gfrak$ and $\gfrak'$ relative to this choice of generators. A field configuration is then simply an infinitely differentiable map $\Phi$ from $\RR^{1,d-1}$ to the bi-adjoint representation of $G \times G'$ on $\gfrak \otimes \gfrak'$. Such a map can be described in terms of complex-valued infinitely differentiable functions $\Phi_{a a'}$ on $\RR^{1,d-1}$ as $\Phi = \Phi_{a a'} T^{a} \otimes T^{\prime a'}$. The action for the theory is thus
\begin{equation}\label{eqn:2.7}
S_{\mathrm{BA}}[\Phi] = \int_{\RR^{1,d-1}}  \ud^d x \left\{ -\frac{1}{2}  \Phi^{a a'} \square \Phi_{a a'} + \frac{1}{3!} f^{abc}f^{\prime a'b'c'} \Phi_{aa'}\Phi_{bb'}\Phi_{cc'}\right\},
\end{equation}
where we have set $\Phi^{aa'} = \kappa^{a b} \kappa^{\prime a' b'} \Phi_{b b'}$, $f^{abc} = \kappa^{cd} f^{ab}_{\phantom{ab}d}$ and $f^{\prime a'b'c'} = \kappa^{\prime c' d'}f^{\prime a'b'}_{\phantom{a'b'}d'}$. The cubic interaction of the ``bi-adjoint scalars'' $\Phi_{aa'}$  manifest the double-copy structure between the ``colour group'' $G$ and the ``dual colour group'' $G'$. 

The equation of motion derived from \eqref{eqn:2.7} has the form
\begin{equation}\label{eqn:2.8}
\square \Phi_{c c'} = \frac{1}{2} f^{ab}_{\phantom{ab}c} f^{\prime a'b'}_{\phantom{a'b'}c'} \Phi_{aa'} \Phi_{b b'}.
\end{equation}
This equation may be put into a more intrinsic form as follows. Denote by $C^{\infty}(\RR^{1,d-1},\gfrak \otimes \gfrak')$ \linebreak the space of infinitely differentiable functions on $\RR^{1,d-1}$ with values in the bi-adjoint representation of $G \times G'$ on $\gfrak \otimes \gfrak'$. On $C^{\infty}(\RR^{1,d-1},\gfrak \otimes \gfrak')$, we can define a binary operation by means of
\begin{align}\label{eqn:2.9}
\begin{split}
\llbracket \Phi,\Psi \rrbracket &= -  \frac{1}{2} \left( \Phi_{a a'} \Psi_{b b'} + \Psi_{a a'} \Phi_{b b'}\right) [T^a,T^b] \otimes [T'^{a'},T'^{b'}]  \\
&=  \frac{1}{2} f^{ab}_{\phantom{ab}c} f'^{a'b'}_{\phantom{a'b'}c'} \left( \Phi_{a a'} \Psi_{b b'} + \Psi_{a a'} \Phi_{b b'}\right) T^c \otimes T'^{c'}.
\end{split}
\end{align}
Then equation \eqref{eqn:2.8} may be written as
\begin{equation}\label{eqn:2.10}
\square \Phi = \frac{1}{2} \llbracket \Phi,\Phi \rrbracket,
\end{equation}
where we have put $\square \Phi = \square \Phi_{a a'} T^a \otimes T'^{a'}$. As the reader may notice, this is a rather difficult and truly nonlinear partial differential equation, whose space of solutions is hard to describe (nonetheless, see \cite{White:2016jzc} and \cite{DeSmet:2017rve} for a number of solutions in which the interaction term is nonzero). Instead, one may try to linearise it by constructing perturbatively a solution in terms of the so-called Berends-Giele double currents. This is the perturbiner expansion. To present it we need some notation.

By a \emph{word} we mean a finite string $I = i_1 i_2 \cdots i_m$ of positive integers $i_1,i_2,\dots,i_m \geq 1$. The word consisting of no symbols is called the \emph{empty word}, written $\varnothing$. Given a word $I = i_1 i_2 \cdots i_m$, we denote by $\bar{I} = i_{m} i_{m-1} \cdots i_1$ its \emph{transpose} and by $\vert I \vert$ its \emph{length} $m$. We further denote by $\mathcal{W}_m$ the set of words of length $m$. If $I$ and $J$ are words, the so is their \emph{concatenation} $IJ$ obtained by juxtaposition, that is, writing $I$ and $J$ after one another. The length of a concatenated word is the sum of the lengths of the concatenatees. We also have the property that $\varnothing I = I \varnothing = I$ for any word $I$.  

By a \emph{shuffle} of two words $I = i_1 i_2 \cdots i_m$ and $J = j_1 j_2 \cdots j_n$, denoted $I \shuffle J$, we understand the formal linear combination 
\begin{equation}
I \shuffle J = \sum_{\sigma \in \mathfrak{S}_{m,n}} I \shuffle_{\sigma} J,
\end{equation}
where the sum is extended to all $(m,n)$-shuffles and where $I \shuffle_{\sigma} J$ is the word resulting by concatenating $I$ and $J$ to get $IJ = i_1 i_2 \cdots i_m j_1 j_2 \cdots j_n = r_1 r_2 \cdots r_{m+n}$ and then permuting letters in such a way to achieve $r_{\sigma(1)} r_{\sigma(2)} \cdots r_{\sigma(m+n)}$. Alternatively, the shuffle operation can be defined inductively by setting
\begin{equation}
\varnothing \shuffle I  = I \shuffle \varnothing = I, \quad i I \shuffle j J = i (I \shuffle j J) + j (i I \shuffle J),
\end{equation}
for any words $I$ and $J$ and for any positive integers $i$ and $j$. For example, $i \shuffle j = ij + ji$ and
$$
i_1 i_2 \shuffle j_1 j_2 = i_1 i_2 j_1 j_2 + j_1 j_2 i_1 i_2 + i_1 j_1 (i_2 \shuffle j_2) + j_1 i_1 (i_2 \shuffle j_2).
$$

We now return to our main focus of formulating the perturbiner expansion of the bi-adjoint scalar theory. Let $\boldsymbol{a} = \left(a_{i}\right)_{i \geq 1}$
 and $\boldsymbol{a}' = \left(a'_{i}\right)_{i \geq 1}$ be, respectively, infinite multisets of ``colour indices'' associated with the Lie algebras $\gfrak$ and $\gfrak'$ (that is, unordered sets of elements of $\{1,2,\dots, \dim \gfrak\}$ and $\{1,2,\dots, \dim \gfrak'\}$, possibly with multiplicity). Let also $\left(k_{i}\right)_{i \geq 1}$ be an infinite set of massless momentum vectors in $\RR^{1,d-1}$. If $I = i_1 i_2 \cdots i_m$ is a word, we put $\boldsymbol{a}_I = ( a_{i_1}, a_{i_2},\dots, a_{i_m})$, $\boldsymbol{a}'_I = ( a'_{i_1}, a'_{i_2},\dots, a'_{i_m})$ and $k_I = k_{i_1} + k_{i_2} + \cdots + k_{i_m}$. The \emph{perturbiner} is defined as a solution to the equation \eqref{eqn:2.8} in the shape of a formal expansion in the noncommutative variables $\ue^{\ui k_i \cdot x} T^{a_i}$ and $T'^{a'_j}$.  More precisely, we take an ansatz providing such a solution of the form
 \begin{align}\label{eqn:2.11}
 \begin{split}
\Phi(x) &= \sum_{m \geq 1} \sum_{I,J \in \mathcal{W}_m} \phi_{I \vert J} \ue^{\ui k_I \cdot x} T^{\boldsymbol{a}_I} \otimes T'^{\boldsymbol{a}'_J}  \\
 &= \sum_{i, j \geq 1} \phi_{i \vert j} \ue^{\ui k_i \cdot x} T^{a_i} \otimes T'^{a'_j} + \sum_{i, j, k, l \geq 1} \phi_{i j \vert kl} \ue^{\ui k_{ij} \cdot x} T^{a_i} T^{a_j} \otimes T'^{a'_k} T'^{a'_l} + \cdots .
 \end{split}
 \end{align}
Here we are employing the notations $T^{a_I} = T^{a_{i_1}}T^{a_{i_2}} \cdots T^{a_{i_m}}$ and $T'^{a'_{J}} = T'^{a'_{j_1}} T'^{a'_{j_2}} \cdots T'^{a'_{j_m}}$ for the products of the Lie algebra generators associated with the colour multi-labels $\boldsymbol{a}_I = ( a_{i_1}, a_{i_2},\dots, a_{i_m})$ and $\boldsymbol{a}'_J = (a'_{j_1},a'_{j_2},\dots, a'_{j_m})$. Also, the perturbiner coefficients $\phi_{I \vert J}$, which are known as the \emph{Berends-Giele double-currents}, are assumed to vanish unless the word $I$ is a permutation of the word $J$. This condition ensures a well-defined multi-particle interpretation. 
 
By inserting the perturbiner expansion \eqref{eqn:2.11} into the equation of motion \eqref{eqn:2.10} and collecting terms of equal number of generators $T^{a_i}$ and $T'^{a'_j}$ on both sides, one obtains a recursion relation for the Berends-Giele double-currents. As found in \cite{Mafra:2016ltu}, this recursion relation takes the form
\begin{equation}\label{eqn:2.12}
\phi_{I \vert J} = \frac{1}{s_I} \sum_{I = KL} \sum_{J = M N} \left( \phi_{K \vert M} \phi_{L \vert N} - \phi_{L \vert M} \phi_{K \vert N} \right),
\end{equation}
where $s_I = k_I^2$ is the \emph{Mandelstam invariant}, and where the notation $\sum_{I = KL}$ and $\sum_{J = MN}$ instructs to sum over deconcatenations of the word $I$ into non-empty words $K$ and $L$ and to independently deconcatenate $J$ in the same manner. In view of the antisymmetry of the right-hand side of \eqref{eqn:2.12} under interchange of the words $K$ and $L$ and $M$ and $N$, one also derives the shuffle constraint $\phi_{I \shuffle J \vert K} = 0$ which, in particular, implies that $\phi_{Ii \vert J} = (-1)^{\vert I \vert} \phi_{i \bar{I} \vert J}$. It is this condition what guarantees that the expression in \eqref{eqn:2.11} takes values in the bi-adjoint representation of $G \times G'$ (which was not clear \emph{a priori}). Let us further remark that one more constraint imposed by the equation of motion \eqref{eqn:2.10} is that the single index double currents $\phi_{i \vert j}$ must be chosen in such a way that the first term in \eqref{eqn:2.11} satisfies the linearised equation $\square \Phi = 0$, which is equivalent to requiring that the momentum vectors $(k_i)_{i \geq 1}$ be lightlike. We may as well normalize and set $\phi_{i \vert j} = \delta_{ij}$. In light of this, we can now appreciate the significance of the perturbiner expansion \eqref{eqn:2.11}: One is trying to write a solution of the nonlinear equation \eqref{eqn:2.10} first as a linear approximation in the variables $\ue^{\ui k_i \cdot x}T^{a_i}$ and $T'^{a'_j}$ and then adding successive nonlinear corrections of higher and higher order in these variables to obtain the complete solution. 

Before closing this subsection, it may be useful to mention how the Berends-Giele double-currents $\phi_{I \vert J}$ are related to tree-level scattering amplitudes of the bi-adjoint scalar theory. As discussed in \cite{Cachazo:2013iea}, the full scattering amplitude of $n$ bi-adjoint scalars can be expanded in the trace decomposition
\begin{align}\label{eq:2.27}
\begin{split}
\mathscr{M}_{n}^{\mathrm{tree}} =  \frac{1}{n} \sum_{i,j \geq 1} \sum_{I,J \in \mathcal{W}_{n-1}} \delta(k_{iI}) m(iI \vert jJ) \tr(T^{\boldsymbol{a}_{iI}}) \tr(T'^{\boldsymbol{a}'_{jJ}}),
\end{split}
\end{align}
where each $m(iI \vert jJ)$ is known as a doubly colour-ordered partial amplitude. It computes the sum of all trivalent scalar diagrams that can be regarded as $\boldsymbol{a}_{iI}$ colour-ordered and $\boldsymbol{a}'_{jJ}$ colour-ordered. In \cite{Mafra:2016ltu}, it was shown that the partial amplitudes can be determined using 
\begin{equation}\label{eq:2.28}
m(iI \vert jJ) = s_{I} \phi_{i \vert j} \phi_{I \vert J}.  
\end{equation}
Thus, knowing all the Berends-Giele double currents gives us all the tree-level scattering amplitudes. As a consequence of the shuffle constraints, the partial amplitudes \eqref{eq:2.28} satisfy the Kleiss-Kuijf relation \cite{Kleiss:1988ne}. We also note that $\mathscr{M}_{2}^{\mathrm{tree}} =0$, because $s_i = 0$ for all $i \geq 1$.

%%%%%%%%%%%%%%%%%%%%%%%%%%%%%%%%%%%%%%%%

\subsection{Yang-Mills theory and its perturbiner expansion}\label{sec:2.4}
In this subsection, we very briefly review the perturbiner expansion for Yang-Mills theory in general spacetime dimensions, following the discussion in \cite{Mafra:2016ltu}. The reader might see \cite{Mizera:2018jbh} for a more thorough treatment. 

Let $G$ be a compact semi-simple Lie group with Lie algebra $\gfrak$. We denote by $T^{a}$ the generators of $\gfrak$, with structure constants $f^{ab}_{\phantom{ab}c}$ satisfying $[T^{a},T^{b}]= \ui f^{ab}_{\phantom{ab}c} T^{c}$. We also let $\kappa^{ab}=\tr (T^{a}T^{b})$ be the components of the Cartan-Killing form on $\gfrak$ with respect to these generators. We consider a Yang-Mills field $A$ as a one-form on $\RR^{1,d-1}$ with values in $\gfrak$. The field strenght $F$ is obtained by taking the covariant exterior derivative of $A$, yielding a two-form on $\RR^{1,d-1}$ with values in $\gfrak$. Using the standard coordinates on $\RR^{1.d-1}$, we can write $A = \ui A_{\mu} \ud x^{\mu}$ and $F =\frac{\ui}{2} F_{\mu\nu}  \ud x^{\mu} \wedge \ud x^{\nu}$ where $
F_{\mu\nu} = \partial_{\mu} A_{\nu} - \partial_{\nu} A_{\mu} + \ui [A_{\mu},A_{\nu}]$. Since $A_{\mu}$ and $F_{\mu\nu}$ are $\gfrak$-valued functions, they may be expanded in terms of the generators $T^{a}$ as $A_{\mu} = A_{\mu a} T^{a}$ and $F_{\mu\nu} = F_{\mu\nu a} T^{a}$. The Yang-Mills action is then
\begin{equation} \label{eqn:2.13}
S_{\mathrm{YM}}[A] = -\frac{1}{4} \int_{\RR^{1,d-1}} \ud^d x\,   F_{\mu\nu}^{a} F^{\mu\nu}_{a},
\end{equation}
where here $F_{\mu\nu}^{a} = \kappa^{a b}F_{\mu\nu b}$. The equation of motion derived from \eqref{eqn:2.13} is
\begin{equation}
\partial_{\mu} F^{\mu\nu} + \ui [A_{\mu},F^{\mu\nu}] = 0, 
\end{equation}
which can be usefully rewritten by imposing the Lorenz gauge condition $\partial_{\mu} A^{\mu} = 0$ as
\begin{equation}\label{eqn:2.14}
\square A^{\nu} = \ui [A_{\mu}, \partial^{\mu} A^{\nu} + F^{\mu \nu}] . 
\end{equation}
In order to define a perturbiner expansion analogous to \eqref{eqn:2.11}, we pick an infinite multiset of ``colour indices'' $\boldsymbol{a} = (a_{i})_{i \geq 1}$ and an infinite set $(k_{i})_{i \geq 1}$ of massless momentum vectors in $\RR^{1,d-1}$. We then look for a solution to the equation \eqref{eqn:2.14} by making the anstaz
\begin{align}\label{eqn:2.15}
\begin{split}
A^{\mu} (x) = \sum_{m \geq 1} \sum_{I \in \mathcal{W}_m} \mathcal{A}^{\mu}_I \ue^{\ui k_{I} \cdot x} T^{\boldsymbol{a}_{I}} = \sum_{i \geq 1} \mathcal{A}^{\mu}_{i} \ue^{\ui k_{i} \cdot x} T^{a_{i}} + \sum_{i,j \geq 1} \mathcal{A}^{\mu}_{ij} \ue^{\ui k_{ij} \cdot x} T^{a_{i}}T^{a_{j}} + \cdots.
\end{split}
\end{align}
Here, as before, we are using the collective notation $T^{a_I} = T^{a_{i_1}} T^{a_{i_2}} \cdots T^{a_{i_m}}$ for each colour multi-label $\boldsymbol{a}_I = (a_{i_1},a_{i_2},\dots,a_{i_m})$. Inserting this expansion back in \eqref{eqn:2.14}, and equating coefficients with the same number of generators $T^{a_i}$ on both sides, gives the following recursion relation \cite{Mafra:2016ltu,Mizera:2018jbh}
\begin{equation}\label{eqn:2.16}
\mathcal{A}^{\mu}_{I} = \frac{1}{s_{I}} \sum_{I = JK} \left\{ (k_{J} \cdot \mathcal{A}_{J}) \mathcal{A}^{\mu}_{K} + \mathcal{A}_{J \nu} \mathcal{F}^{\mu \nu}_{K}  - (k_{K} \cdot \mathcal{A}_{K}) \mathcal{A}^{\mu}_{J} - \mathcal{A}_{K \nu} \mathcal{F}^{\mu \nu}_{J}  \right\}, 
\end{equation}
where we have introduced the quantity
\begin{equation}
\mathcal{F}^{\mu\nu}_{I} = k^{\mu}_{I} \mathcal{A}^{\nu}_{I} - k^{\nu}_{I} \mathcal{A}^{\mu}_{I} - \sum_{I= JK} \left( \mathcal{A}^{\mu}_{J} \mathcal{A}^{\nu}_{K} - \mathcal{A}^{\mu}_{K} \mathcal{A}^{\nu}_{J} \right). 
\end{equation}
Also, it is straightforward to derive the Berends-Giele symmetry associated to a shuffle constraint $\mathcal{A}^{\mu}_{I \shuffle J} = 0$ or, equivalently, $\mathcal{A}^{\mu}_{IiJ} = (-1)^{\vert I \vert} \mathcal{A}^{\mu}_{i(\bar{I} \shuffle J)}$. This guarantees that the ansatz \eqref{eqn:2.15} takes values in the Lie algebra $\gfrak$. In addition, the single labeled coefficients $\mathcal{A}^{\mu}_{i}$ are to be chosen so that that the first term in \eqref{eqn:2.15} satisfies the linearised equation $\square A^{\mu} = 0$, which is equivalent to imposing that the momentum vectors $(k_i)_{i \geq 1}$ be lightlike. By bringing back to mind the Lorentz gauge condition, we are lead to conclude that $\mathcal{A}^{\mu}_{i} = \varepsilon^{\mu}_{i}$ are polarisation vectors that satisfy the transversality condition $k_i \cdot \varepsilon _i = 0$.  

It remains to say a word about scattering amplitudes in Yang-Mills theory. At tree level, the full scattering amplitude of $n$ gluons can be decomposed as
\begin{equation}\label{eq:2.35}
\mathscr{A}_n^{\mathrm{tree}} = \frac{1}{n} \sum_{i \geq 1} \sum_{I \in \mathcal{W}_{n-1}} \delta(k_{iI}) A(iI) \tr (T^{\boldsymbol{a}_{iI}}),
\end{equation}
where $A(iI)$ is the colour-ordered partial amplitude, which contains all the kinematic information. As shown in \cite{Berends:1987me}, it is determined by the Berends-Giele currents $\mathcal{A}_{I}^{\mu}$ through the formula
\begin{equation}\label{eq:2.36}
A(iI) = s_{I} \mathcal{A}_{i} \cdot \mathcal{A}_{I}. 
\end{equation}
Also, by virtue of the shuffle constraints, this partial amplitude satisfies the Kleiss-Kuijf relation \cite{Kleiss:1988ne}. Finally we note that $\mathscr{A}_2^{\mathrm{tree}} = 0$, since $s_i = 0$ for all $i \geq 1$. 

%%%%%%%%%%%%%%%%%%%%%%%%%%%%%%%%%%%%%%%%%%%%%%%%%%%%%%%%%%%%%%%%%%%%%

\section{Perturbiner expansions and minimal models}\label{sec:3}
Having reviewed the requisite mathematical machinery, let us get to the problem at hand, namely to show how the perturbiner expansion for the bi-adjoint scalar and Yang-Mills theories is obtained in the transition to the minimal model of their corresponding $L_{\infty}$-algebras. The results for the Yang-Mills theory are intimately related to those recently obtained by Macrelli, S\"{a}mann and Wolf in \cite{Macrelli:2019afx}, though we remark that these authors did not work in the perturbiner framework. 

\subsection{The bi-adjoint scalar DG Lie algebra}\label{sec:3.1}
This subsection will describe the DG Lie algebra associated to the bi-adjoint scalar theory.  Our notation here is the same as in \S \ref{sec:2.3}. 

As discussed more fully in \cite{Jurco:2018sby}, the classical Batalin-Vilkovisky formalism assigns to any field theory it can treat a cyclic $L_{\infty}$-algebra encoding its symmetries, field content, equations of motion, and Noether currents. In the case of bi-adjoint scalar theory, since there is no gauge symmetry to be fixed, the associated cyclic $L_{\infty}$-algebra is actually a cyclic DG Lie algebra which we call $\LBA$. As a cochain complex, $\LBA$ is
$$
C^{\infty}(\RR^{1,d-1},\gfrak \otimes \gfrak')[-1] \xlongrightarrow{- \square}  C^{\infty}(\RR^{1,d-1},\gfrak \otimes \gfrak')[-2]. 
$$
Thus, $C^{\infty}(\RR^{1,d-1},\gfrak \otimes \gfrak')$ is situated in degrees $1$ and $2$, and the differential $l_1$ is the negative of the d'Alembertian operator $\square$ acting on $C^{\infty}(\RR^{1,d-1},\gfrak \otimes \gfrak')$. The binary operation
$$
l_2 \colon C^{\infty}(\RR^{1,d-1},\gfrak \otimes \gfrak')[-1] \otimes C^{\infty}(\RR^{1,d-1},\gfrak \otimes \gfrak')[-1] \lto C^{\infty}(\RR^{1,d-1},\gfrak \otimes \gfrak')[-2],
$$
is defined by 
\begin{equation}
l_2 (\Phi,\Psi) = \llbracket \Phi,\Psi \rrbracket. 
\end{equation}
Recalling the definition \eqref{eqn:2.9}, this is evidently skew-symmetric and since the graded Jacobi identity is trivially satisfied, it turns 
\begin{equation}\label{eqn:3.1}
\LBA = C^{\infty}(\RR^{1,d-1},\gfrak \otimes \gfrak')[-1] \oplus C^{\infty}(\RR^{1,d-1},\gfrak \otimes \gfrak')[-2]
\end{equation}
into a DG Lie algebra. This DG Lie algebra can be made cyclic by setting 
\begin{equation}\label{eqn:3.2}
\langle \Phi,\Psi \rangle = \int_{\RR^{1,d-1}} \ud^{d} x \, \Phi^{aa'}\Psi_{aa'},
\end{equation}
where, as before, $\Phi = \Phi_{aa'} T^{a} \otimes T'^{a'}$, $\Psi = \Psi_{aa'} T^{a} \otimes T'^{a'}$ and $\Phi^{aa'} = \kappa^{ab}\kappa'^{a'b'}\Phi_{bb'}$. With this choice, we find that
$$
\frac{1}{2} \langle \Phi,l_1(\Phi) \rangle = - \frac{1}{2} \int_{\RR^{1,d-1}} \ud^{d} x \, \Phi^{aa'} \square \Phi_{aa'},
$$
and
$$
\frac{1}{3!} \langle \Phi,l_2(\Phi,\Phi) \rangle = \frac{1}{3!} \int_{\RR^{1,d-1}} \ud^{d} x \, f^{abc}f'^{abc} \Phi_{aa'}\Phi_{bb'}\Phi_{cc'}.
$$
Thus, the homotopy Maurer-Cartan action for the DG Lie algebra \eqref{eqn:3.1} is
\begin{equation*}
S_{\mathrm{MC}}[\Phi] = \int_{\RR^{1,d-1}} \ud^{d} x \left\{- \frac{1}{2} \Phi^{aa'} \square \Phi_{aa'} + \frac{1}{3!} f^{abc}f'^{abc} \Phi_{aa'}\Phi_{bb'}\Phi_{cc'} \right\},
\end{equation*}
which of course coincides with the action for the bi-adjoint scalar theory \eqref{eqn:2.7}. It is, perhaps, worth mentioning that we have omitted here all the technicalities of choosing the appropriate fall-off conditions in our function spaces. This can be done using the ideas considered in \S 3 of \cite{Macrelli:2019afx}. 

Now, in order to deal with the perturbiner expansion for the bi-adjoint scalar theory, we need to change slightly the definition of the DG Lie algebra \eqref{eqn:3.1}. So, let us fix infinite multisets of ``colour indices'' $\boldsymbol{a} = (a_{i})_{i \geq 1}$ and $\boldsymbol{a}' = (a'_{i})_{i \geq 1}$ associated to the Lie algebras $\gfrak$ and $\gfrak'$, respectively, was well as an infinite set $(k_i)_{i \geq 1}$ of massless momentum vectors in $\RR^{1,d-1}$. Denote by $\mathscr{E}(\RR^{1,d-1},\gfrak \otimes \gfrak')$ the space of formal series of the form
\begin{align}\label{eqn:3.3}
\begin{split}
\Phi(x) &= \sum_{m \geq 1} \sum_{I,J \in \mathcal{W}_m} \phi_{I \vert J} \ue^{\ui k_{I} \cdot x} T^{\boldsymbol{a}_I} \otimes T'^{\boldsymbol{a}'_J} \\
&= \sum_{i ,j \geq 1} \phi_{i \vert j} \ue^{\ui k_{i} \cdot x} T^{a_i} \otimes T'^{a'_j} + \sum_{i,j,k,l \geq 1} \phi_{ij \vert kl} \ue^{\ui k_{ij} \cdot x} T^{a_i}T^{a_j} \otimes T'^{a'_k}T'^{a'_l} + \cdots ,    
\end{split}
\end{align}
where the coefficients $\phi_{I \vert J}$ are supposed to vanish unless the word $I$ is a permutation of the word $J$. In line with the the terminology used in \cite{Mizera:2018jbh}, we may refer to the elements of $\mathscr{E}(\RR^{1,d-1},\gfrak \otimes \gfrak')$ as \emph{colour-stripped perturbiner ansatzs}. 

To go further, we must say a bit about how to extend the d'Alembertian operator $\square$ and the binary operation $\llbracket , \rrbracket$ to $\mathscr{E}(\RR^{1,d-1},\gfrak \otimes \gfrak')$. This will be provided by what in reference \cite{Mizera:2018jbh} is called the \emph{colour-dressed} version of the perturbiner ansatz. We therefore introduce, for each ordered sequence of positive integers $i_1 < i_2 < \cdots < i_m$, the notations
\begin{align}
\begin{split}
f^{i_1 i_2 \cdots i_m}_{\phantom{i_1 i_2 \cdots i_m} a} &= f^{a_{i_1} a_{i_2}}_{\phantom{a_{i_1} a_{i_2}} b} f^{b a_{i_3}}_{\phantom{b a_{i_3}} c} \cdots f^{d a_{i_{m-1}}}_{\phantom{d a_{i_{m-1}}} e} f^{e a_{i_{m}}}_{\phantom{e a_{i_{m}}} a},  \\
f'^{i_1 i_2 \cdots i_m}_{\phantom{i_1 i_2 \cdots i_m} a'} &= f'^{a'_{i_1} a'_{i_2}}_{\phantom{a'_{i_1} a'_{i_2}} b'} f'^{b' a'_{i_3}}_{\phantom{b' a'_{i_3}} c'} \cdots f'^{d' a'_{i_{m-1}}}_{\phantom{d' a'_{i_{m-1}}} e'} f'^{e' a'_{i_{m}}}_{\phantom{e' a'_{i_{m}}} a'},  
\end{split}
\end{align}
and define
\begin{equation} 
\phi_{aa' \vert i_{1}i_{2}\cdots i_{m}} = \sum_{\sigma,\tau \in \mathfrak{S}_{m-1}} f^{i_1 i_{\sigma(2)} \cdots i_{\sigma(m)}}_{\phantom{i_1  i_{\sigma(2)} \cdots i_{\sigma(m)}} a} f'^{i_1 i_{\tau(2)} \cdots i_{\tau(m)}}_{\phantom{i_1  i_{\tau(2)} \cdots i_{\tau(m)}} a'} \phi_{i_1 i_{\sigma(2)} \cdots i_{\sigma(m)} \vert i_1 i_{\tau(2)} \cdots i_{\tau(m)}}.
\end{equation}
With the help of the latter, equation \eqref{eqn:3.3} becomes simply $\Phi(x) = \Phi_{aa'}(x) T^{a} \otimes T'^{a'}$, 
where the coefficients $\Phi_{aa'}(x)$ are formal series of the form
\begin{align}
\begin{split}
\Phi_{aa'}(x) &= \sum_{m \geq 1}  \sum_{I \in \mathcal{OW}_m }\phi_{aa' \vert I} \ue^{\ui k_{I} \cdot x}  \\
 &= \sum_{i } \phi_{aa' \vert i} \ue^{\ui k_i \cdot x} +  \sum_{i < j} \phi_{aa' \vert ij} \ue^{\ui k_{ij} \cdot x} + \sum_{i < j < k } \phi_{aa' \vert ijk} \ue^{\ui k_{ijk} \cdot x} + \cdots.
\end{split}
\end{align}
Here $\mathcal{OW}_m$ denotes the set of words $I = i_1 i_2 \cdots i_m$ of length $m$ with $i_1 < i_2 < \cdots < i_m$. This enables us to define the d'Alembertian operator acting on the space $\mathscr{E}(\RR^{1,d-1},\gfrak \otimes \gfrak')$ as $\square \Phi(x) = \square \Phi_{aa'}(x) T^{a} \otimes T'^{a'}$, where the $\square \Phi_{aa'}(x)$ are given by
\begin{align}
\begin{split}
\square \Phi_{aa'}(x) &= -\sum_{m \geq 1}  \sum_{I \in \mathcal{OW}_m }s_{I} \phi_{aa' \vert I} \ue^{\ui k_{I} \cdot x}  \\
 &= -\sum_{i } s_i \phi_{aa' \vert i} \ue^{\ui k_i \cdot x} -  \sum_{i < j} s_{ij} \phi_{aa' \vert ij} \ue^{\ui k_{ij} \cdot x} - \sum_{i < j < k }s_{ijk} \phi_{aa' \vert ijk} \ue^{\ui k_{ijk} \cdot x} + \cdots.
\end{split}
\end{align}
It also allows us to define a bracket operation on $\mathscr{E}(\RR^{1,d-1},\gfrak \otimes \gfrak')$ by the same expression as that of \eqref{eqn:2.9}. A direct calculation then shows $\mathscr{E}(\RR^{1,d-1},\gfrak \otimes \gfrak')$ is closed under this bracket (see the argument presented in \S 2.2 of \cite{Mizera:2018jbh}).  

In light of the above discussion, the cochain complex underlying the cyclic DG Lie algebra $\LBA$ that encodes the perturbiner expansion for the bi-adjoint scalar theory is
\begin{equation}\label{eqn:3.4}
\mathscr{E}(\RR^{1,d-1},\gfrak \otimes \gfrak')[-1] \xlongrightarrow{- \square}  \mathscr{E}(\RR^{1,d-1},\gfrak \otimes \gfrak')[-2]. 
\end{equation}
Thus, as before, the differential $l_1$ is the negative of the d'Alembertian operator $\square$ acting on $\mathscr{E}(\RR^{1,d-1},\gfrak \otimes \gfrak')$. As for the binary operation
$$
l_2 \colon \mathscr{E}(\RR^{1,d-1},\gfrak \otimes \gfrak')[-1] \otimes \mathscr{E}(\RR^{1,d-1},\gfrak \otimes \gfrak')[-1] \lto \mathscr{E}(\RR^{1,d-1},\gfrak \otimes \gfrak')[-2],
$$
it is again determined by the bracket operation $\llbracket , \rrbracket$. Finally, $\LBA$ can be augmented to a cyclic DG Lie algebra by means of the symmetric pairing \eqref{eqn:3.2}. 

%%%%%%%%%%%%%%%%%%%%%%%%%%%%%%%%%%%%%%%%%%%%%%%%%%%%%%%

\subsection{The perturbiner expansion for bi-adjoint scalar theory revisited} \label{sec:3.2}
In this subsection, we will show that the determination of the perturbiner expansion for the bi-adjoint scalar theory can be reduced 
to the construction of a minimal model for the bi-adjoint DG Lie algebra $\LBA$ introduced above. We shall adhere to the terminology and notation employed in \S \ref{sec:2.1}. 

To begin with, using the defining cochain complex \eqref{eqn:3.4}, we see that the cohomology of $\LBA$ is concentrated in degrees $1$ and $2$. It is given by the solution space $H^{1}(\LBA) = \ker \left( l_1\right)$ of the linearaised equation $\square \Phi = 0$ and the space $H^{2}(\LBA) = \mathscr{E}(\RR^{1,d-1},\gfrak \otimes \gfrak')/ \im \left( l_1 \right)$ of linear on-shell colour-stripped perturbiner ansatzs. It follows that the cochain complex underlying the cohomology $H^{\sbullet}(\LBA)$ of $\LBA$ is 
$$
 \ker \left( l_1\right)[-1] \xlongrightarrow{0} \mathscr{E}(\RR^{1,d-1},\gfrak \otimes \gfrak')/ \im \left( l_1 \right)[-2].
$$
On the other hand, in order to construct the minimal $L_{\infty}$-structure on $H^{\sbullet}(\LBA)$, we must define a projection $p \colon \LBA \to H^{\sbullet}(\LBA)$ and a contracting homotopy $h \colon \LBA \to \LBA$. To this end, we consider the Feynman propagator $G^{\mathrm{F}}$ defined on the space of complex-valued infinitely differentiable functions on $\RR^{1,d-1}$. We do not need here the precise formula for $G^{\mathrm{F}}$, but merely the fact that it is linear and satisfies
\begin{equation}\label{eqn:3.5}
G^{\mathrm{F}} (\ue^{\ui k \cdot x}) = \frac{\ue^{\ui k \cdot x}}{k^2}, 
\end{equation}
as long as $k$ is not lightlike. Just as we did for the d'Alembertian operator $\square$, we extend $G^{\mathrm{F}}$ to all of $\mathscr{E}(\RR^{1,d-1},\gfrak \otimes \gfrak')$  so that we obtain a linear operator $G^{\mathrm{F}} \colon \mathscr{E}(\RR^{1,d-1},\gfrak \otimes \gfrak') \to \mathscr{E}(\RR^{1,d-1},\gfrak \otimes \gfrak')$ satisfying 
\begin{equation}
l_1 \circ G^{\mathrm{F}} = -\square \circ  G^{\mathrm{F}}  = \id_{\mathscr{E}(\RR^{1,d-1},\gfrak \otimes \gfrak')}.  
\end{equation}
With the help of $G^{\mathrm{F}}$, we may define the projection $p^{(1)} \colon \mathscr{E}(\RR^{1,d-1},\gfrak \otimes \gfrak') \to \ker \left( l_1 \right)$ by 
\begin{equation}
p^{(1)} = \id_{\mathscr{E}(\RR^{1,d-1},\gfrak \otimes \gfrak')} - \GF\circ l_1.
\end{equation}
As for the other projection $p^{(2)} \colon \mathscr{E}(\RR^{1,d-1},\gfrak \otimes \gfrak') \to \mathscr{E}(\RR^{1,d-1},\gfrak \otimes \gfrak')/ \im \left( l_1 \right)$ we simply take the quotient map. Thus, it only remains to define a contracting homotopy $h \colon \LBA \to \LBA$. This may be visualised by the left-down pointing arrows in the diagram
$$
\xymatrix@R=10ex@C=10ex{0 \ar[r]^-{0} \ar[d]_-{0} & \mathscr{E}(\RR^{1,d-1},\gfrak \otimes \gfrak') \ar[r]^-{l_1} \ar[ld]^-{h^{(1)}= 0}  \ar@<0.15cm>[d]^-{i^{(1)} \circ p^{(1)}} \ar@<-0.15cm>[d]_-{\id} & \mathscr{E}(\RR^{1,d-1},\gfrak \otimes \gfrak') \ar[r]^-{0} \ar[ld]^-{h^{(2)}} \ar@<0.15cm>[d]^-{i^{(2)} \circ p^{(2)}} \ar@<-0.15cm>[d]_-{\id} & 0 \ar[d]^-{0} \ar[ld]^-{h^{(3)}=0} \\
0 \ar[r]_-{0} & \mathscr{E}(\RR^{1,d-1},\gfrak \otimes \gfrak') \ar[r]_-{l_1} & \mathscr{E}(\RR^{1,d-1},\gfrak \otimes \gfrak') \ar[r]_-{0}& 0, }
$$
so in the present case is simply given by a single linear operator $h^{(2)} \colon \mathscr{E}(\RR^{1,d-1},\gfrak \otimes \gfrak') \to \mathscr{E}(\RR^{1,d-1},\gfrak \otimes \gfrak')$. The condition $\id_{\LBA} - i \circ p = l_1 \circ h + h \circ l_1$ is therefore equivalent to the two equalities
\begin{equation}
 h^{(2)} \circ l_1 + i^{(1)} \circ p^{(1)} = \id_{\mathscr{E}(\RR^{1,d-1},\gfrak \otimes \gfrak')}, \qquad l_1 \circ h^{(2)} + i^{(2)} \circ p^{(2)} =\id_{\mathscr{E}(\RR^{1,d-1},\gfrak \otimes \gfrak')} .
\end{equation}
It is also obvious that the operator $G^{\mathrm{F}}$ automatically satisfies both relations. Hence, we must take $h^{(2)} = G^{\mathrm{F}}$. 

We can now readily compute the minimal $L_{\infty}$-structure on $H^{\sbullet}(\LBA)$ using the formulas \eqref{eqn:2.5} and \eqref{eqn:2.6}. To keep the discussion and derivations as simple as possible, we will limit ourselves to the case where $\Phi'_1,\dots,\Phi'_n \in H^{1}(\LBA) = \ker \left( l_1\right)$ so that $\chi(\sigma; \Phi'_1,\dots,\Phi'_n) = 1$ for any $\sigma \in \mathfrak{S}_{i, n-i}$. Thus $f_1$ is simply the natural embedding $i^{(1)} \colon \ker \left( l_1 \right) \to \mathscr{E}(\RR^{1,d-1},\gfrak \otimes \gfrak')$, while the action of $f_n$ on $\Phi'_1,\dots,\Phi'_n$ is determined recursively by the equation
\begin{align}\label{eqn:3.6}
\begin{split}
f_n (\Phi'_1,\dots,\Phi'_n) &=  -  \tfrac{1}{2} \sum_{i =1}^{n-1} \sum_{\sigma \in \mathfrak{S}_{i, n-i}} (\GF \! \circ\hspace{0.1ex} l_2) (f_i (\Phi'_{\sigma (1)} ,\dots, \Phi'_{\sigma(i)} ), f_{n-i}(\Phi'_{\sigma (i+1)} ,\dots, \Phi'_{\sigma(n)}))  \\
&= -  \tfrac{1}{2} \sum_{i =1}^{n-1} \sum_{\sigma \in \mathfrak{S}_{i, n-i}} \GF \left( \big\llbracket f_i (\Phi'_{\sigma (1)} ,\dots, \Phi'_{\sigma(i)} ), f_{n-i}(\Phi'_{\sigma (i+1)} ,\dots, \Phi'_{\sigma(n)}) \big\rrbracket \right).
\end{split}
\end{align}
In addition, the higher order bracket among $\Phi'_1,\dots,\Phi'_n$ is given by
\begin{align}\label{eqn:3.7}
\begin{split}
l'_n (\Phi'_1,\dots,\Phi'_n) &= \tfrac{1}{2} \sum_{i =1}^{n-1} \sum_{\sigma \in \mathfrak{S}_{i, n-i}} (p^{(2)} \circ l_2) (f_i (\Phi'_{\sigma (1)} ,\dots, \Phi'_{\sigma(i)} ), f_{n-i}(\Phi'_{\sigma (i+1)} ,\dots, \Phi'_{\sigma(n)})) \\
&=\tfrac{1}{2} \sum_{i =1}^{n-1} \sum_{\sigma \in \mathfrak{S}_{i, n-i}} p^{(2)} \left( \big\llbracket f_i (\Phi'_{\sigma (1)} ,\dots, \Phi'_{\sigma(i)} ), f_{n-i}(\Phi'_{\sigma (i+1)} ,\dots, \Phi'_{\sigma(n)}) \big\rrbracket \right).
\end{split}
\end{align}
We also have at our disposal the constraints imposed by \eqref{eqn:2.2} and \eqref{eqn:2.3}. 

With all this in place, we are now in a position to discuss how the perturbiner expansion for the bi-adjoint scalar theory arises from the minimal $L_{\infty}$-structure on $H^{\sbullet}(\LBA)$. We start with a Maurer-Cartan element $\Phi' \in H^{1}(\LBA) = \ker \left( l_1 \right)$ of the form
\begin{equation}\label{eqn:3.8}
\Phi' = \sum_{i,j \geq 1} \phi_{i \vert j} \ue^{\ui k_{i} \cdot x} T^{a_i} \otimes T'^{a'_j}.
\end{equation}
This is the simplest multi-particle solution to the linearised equation and can be thought of as a plane-wave superposition of all the particles. Our claim is that the perturbiner expansion is simply given by
\begin{equation}\label{eqn:3.9}
\Phi = \sum_{n \geq 1} \frac{1}{n!} f_{n}(\Phi',\dots,\Phi').
\end{equation}
In other words, one knows the perturbiner expansion as soon as one knows the $L_{\infty}$-quasi-isomorphism between $H^{\sbullet}(\LBA)$ and $\LBA$. To substantiate the claim, each term contributing to the overall sum in \eqref{eqn:3.9} needs to be calculated explicitly. 

For this purpose, observe that the inner sum in \eqref{eqn:3.6} has $\binom{n}{i}$ summands, so that
\begin{equation} \label{eqn:3.10}
f_{n}(\Phi',\dots,\Phi')= -  \tfrac{1}{2} \sum_{i =1}^{n-1} \binom{n}{i} \GF \left( \big\llbracket f_i (\Phi',\dots, \Phi' ), f_{n-i}(\Phi' ,\dots, \Phi') \big\rrbracket \right).
\end{equation}
It is instructive to examine this expression at the first few values of $n$. For $n = 2$, the application of \eqref{eqn:3.9} trivially leads to
\begin{align*}
f_2 (\Phi',\Phi') = -\tfrac{1}{2} \binom{2}{1}  \GF \left( \llbracket f_1(\Phi'),f_1(\Phi') \rrbracket \right) = - \GF \left( \llbracket \Phi',\Phi' \rrbracket \right). 
\end{align*}
Using equations \eqref{eqn:2.10} and \eqref{eqn:3.5}, this is
\begin{align*}
\begin{split}
f_2 (\Phi',\Phi') &= \GF \left( \sum_{i, j, k, l \geq 1} \phi_{i \vert k} \phi_{j \vert l} \ue^{\ui k_{ij} \cdot x} [T^{a_i}, T^{a_j}] \otimes [T'^{a'_{k}}, T'^{a'_{l}}] \right)\\
&=  \GF \left( \sum_{i, j, k, l \geq 1} 2 \left(\phi_{i \vert k} \phi_{j \vert l} - \phi_{i \vert l} \phi_{j \vert k}\right) \ue^{\ui k_{ij} \cdot x} T^{a_i} T^{a_j} \otimes T'^{a'_{k}} T'^{a'_{l}} \right) \\
&=2! \sum_{i, j, k, l \geq 1} \frac{1}{s_{ij}} \left( \phi_{i \vert k} \phi_{j \vert l} - \phi_{i \vert l} \phi_{j \vert k} \right) \ue^{\ui k_{ij} \cdot x} T^{a_i} T^{a_j} \otimes T'^{a'_{k}} T'^{a'_{l}},
\end{split}
\end{align*}
where in the second equality, we have expanded the commutators, reorder the terms and rename the indices. We may then define new coefficients
\begin{equation}\label{eqn:3.11}
\phi_{ij \vert kl} = \frac{1}{s_{ij}} \left( \phi_{i \vert k} \phi_{j \vert l} - \phi_{i \vert l} \phi_{j \vert k} \right),
\end{equation}
and write
\begin{equation}\label{eqn:3.12}
f_2 (\Phi',\Phi') = 2! \sum_{i, j, k, l \geq 1} \phi_{ij \vert kl}  \ue^{\ui k_{ij} \cdot x} T^{a_i} T^{a_j} \otimes T'^{a'_{k}} T'^{a'_{l}}.
\end{equation}
For $n = 3$, taking into account the skew-symmetry of $\llbracket , \rrbracket$, we find
\begin{align*}
\begin{split}
f_3 (\Phi',\Phi',\Phi') &= -\tfrac{1}{2}\binom{3}{1} \GF \left( \llbracket f_1(\Phi'),f_2(\Phi',\Phi') \rrbracket \right) -\tfrac{1}{2}\binom{3}{2} \GF \left( \llbracket f_2(\Phi',\Phi'),f_1(\Phi') \rrbracket \right) \\
&= - 3 \GF \left( \llbracket \Phi',f_2(\Phi',\Phi') \rrbracket \right).
\end{split}
\end{align*}
From equations \eqref{eqn:2.10}, \eqref{eqn:3.5} and \eqref{eqn:3.12}, we have then
\begin{align*}
\begin{split}
f_3 (\Phi',\Phi',\Phi') &= 3! \GF \left( \sum_{i,j,k,l,m,n \geq 1} \phi_{i \vert l} \phi_{jk \vert mn} \ue^{\ui k_{ijk} \cdot x} [T^{a_i},T^{a_j}T^{a_k}] \otimes [T'^{a'_l},T'^{a'_m}T'^{a'_n}]\right) \\
&=3! \GF \left( \sum_{i,j,k,l,m,n \geq 1} \left(  \phi_{i \vert l} \phi_{jk \vert mn} + \phi_{ij \vert lm} \phi_{k \vert n} - \phi_{i \vert n} \phi_{jk \vert lm} - \phi_{ij \vert mn} \phi_{k \vert l} \right) \right. \\
& \quad \left. \phantom{\sum_{i,j,k,l,m,n \geq 1}} \qquad \quad\quad \qquad \qquad \qquad \times \ue^{\ui k_{ijk} \cdot x} T^{a_i}T^{a_j}T^{a_k} \otimes T'^{a'_l}T'^{a'_m}T'^{a'_n}\right) \\
&=3!  \sum_{i,j,k,l,m,n \geq 1} \frac{1}{s_{ijk}}\left(  \phi_{i \vert l} \phi_{jk \vert mn} + \phi_{ij \vert lm} \phi_{k \vert n} - \phi_{i \vert n} \phi_{jk \vert lm} - \phi_{ij \vert mn} \phi_{k \vert l} \right)  \\
& \quad  \phantom{\sum_{i,j,k,l,m,n \geq 1}} \qquad \quad\quad \qquad \qquad \qquad \hspace{0.6ex} \times \ue^{\ui k_{ijk} \cdot x} T^{a_i}T^{a_j}T^{a_k} \otimes T'^{a'_l}T'^{a'_m}T'^{a'_n},
\end{split}
\end{align*}
where again in the second line, we have expanded the commutators, reorder the terms and rename the indices. We can therefore define new coefficients
\begin{equation}\label{eqn:3.13}
\phi_{ijk \vert lmn} = \frac{1}{s_{ijk}}\left(  \phi_{i \vert l} \phi_{jk \vert mn} + \phi_{ij \vert lm} \phi_{k \vert n} - \phi_{i \vert n} \phi_{jk \vert lm} - \phi_{ij \vert mn} \phi_{k \vert l} \right),
\end{equation}
so that the factor of the third summand in \eqref{eqn:3.10} takes the form
\begin{equation}\label{eqn:3.14}
f_3 (\Phi',\Phi',\Phi') = 3! \sum_{i,j,k,l,m,n \geq 1} \phi_{ijk \vert lmn} \ue^{\ui k_{ijk} \cdot x} T^{a_i}T^{a_j}T^{a_k} \otimes T'^{a'_l}T'^{a'_m}T'^{a'_n}.
\end{equation}
For $n = 4$, equation \eqref{eqn:3.9} reads
\begin{align*}
\begin{split}
f_4 (\Phi',\Phi',\Phi',\Phi') &= -\tfrac{1}{2} \binom{4}{1} \GF \left( \llbracket f_1(\Phi'),f_3(\Phi',\Phi',\Phi') \rrbracket \right) - \tfrac{1}{2} \binom{4}{2} \GF \left( \llbracket f_2(\Phi',\Phi'),f_2(\Phi',\Phi') \rrbracket \right) \\
&\phantom{=}\hspace{0.7ex} -  \tfrac{1}{2} \binom{4}{3} \GF \left( \llbracket f_3(\Phi',\Phi',\Phi'),f_1(\Phi') \rrbracket \right) \\
&= -4 \GF \left( \llbracket f_1(\Phi'),f_3(\Phi',\Phi',\Phi') \rrbracket \right) - \tfrac{1}{2} \binom{4}{2} \GF \left( \llbracket f_2(\Phi',\Phi'),f_2(\Phi',\Phi') \rrbracket \right).
\end{split}
\end{align*}
The first term on the right-hand side will involve commutators of the type $[T^{a_i}, T^{a_j}T^{a_k}T^{a_l}]$ and $[T'^{a'_m}, T'^{a'_n}T'^{a'_p}T'^{a'_q}]$. Thus, using \eqref{eqn:3.14}, and proceeding as in the previous calculations, it is found that this term equals
\begin{align*}
4! \sum_{i,j,k,l,m,n,p,q \geq 1} \frac{1}{s_{ijkl}} &\left(\phi_{i \vert m} \phi_{jkl \vert npq} + \phi_{ijk \vert mnp} \phi_{l \vert q} - \phi_{i \vert q} \phi_{jkl \vert mnp} -  \phi_{ijk \vert npq} \phi_{l \vert m} \right) \\
&\qquad \qquad \qquad\qquad \times \ue^{\ui k_{ijkl} \cdot x} T^{a_i}T^{a_j}T^{a_k}T^{a_l} \otimes T'^{a'_m}T'^{a'_n}T'^{a'_p}T'^{a'_q}. 
\end{align*}
Likewise, the second term will involve commutators of the type $[T^{a_i}T^{a_j}, T^{a_k}T^{a_l}]$ and \linebreak $[T'^{a'_m}T'^{a'_n}, T'^{a'_p}T'^{a'_q}]$ and is seen to be equal to
\begin{align*}
4! \sum_{i,j,k,l,m,n,p,q \geq 1} \frac{1}{s_{ijkl}} \left( \phi_{ij \vert mn} \phi_{kl \vert pq} - \phi_{ij \vert pq}\phi_{kl \vert mn} \right) \ue^{\ui k_{ijkl} \cdot x} T^{a_i}T^{a_j}T^{a_k}T^{a_l} \otimes T'^{a'_m}T'^{a'_n}T'^{a'_p}T'^{a'_q}. 
\end{align*}
The factor of the fourth summand in \eqref{eqn:3.10} then becomes
\begin{align}\label{eqn:3.15}
\begin{split}
f_4 (\Phi',\Phi',\Phi',\Phi') = 4! \sum_{i,j,k,l,m,n,p,q \geq 1} \phi_{ijkl \vert mnpq} \ue^{\ui k_{ijkl} \cdot x} T^{a_i}T^{a_j}T^{a_k}T^{a_l} \otimes T'^{a'_m}T'^{a'_n}T'^{a'_p}T'^{a'_q},
\end{split}
\end{align}
where we have now introduced the coefficients
\begin{align}\label{eqn:3.16}
\begin{split}
\phi_{ijkl \vert mnpq} = \frac{1}{s_{ijkl}} (\phi_{i \vert m} \phi_{jkl \vert npq} + \phi_{ijk \vert mnp} \phi_{l \vert q} &- \phi_{i \vert q} \phi_{jkl \vert mnp} -  \phi_{ijk \vert npq} \phi_{l \vert m}  \\
 &\qquad \quad +  \phi_{ij \vert mn} \phi_{kl \vert pq} - \phi_{ij \vert pq}\phi_{kl \vert mn} ).
\end{split}
\end{align}
Comparing equations in \eqref{eqn:3.11}, \eqref{eqn:3.13} and \eqref{eqn:3.16}, we see a pattern emerging. Using the same manipulations that we have been using, we can generate an equation similar to \eqref{eqn:3.12}, \eqref{eqn:3.14} and \eqref{eqn:3.15} for arbitrary $n$. This equation is
\begin{equation}\label{eqn:3.17}
f_n (\Phi',\dots,\Phi') = n! \sum_{I, J \in \mathcal{W}_n} \phi_{I \vert J} \ue^{\ui k_I \cdot x} T^{\boldsymbol{a}_I} \otimes T'^{\boldsymbol{a}'_J}, 
\end{equation}
where the coefficients $\phi_{I \vert J}$ are determined from the recursion relation \eqref{eqn:2.12} for the Berends-Giele double currents. The proof of \eqref{eqn:3.17} is by mathematical induction. In fact, one just have to notice that the $i$th term on the right-hand side of \eqref{eqn:3.10} involves commutators of the type $[T^{\boldsymbol{a}_{K}},T^{\boldsymbol{a}_{M}}]$ and $[T^{\boldsymbol{a}_{L}},T^{\boldsymbol{a}_{N}}]$, where $K$ and $L$ are words of length $i$ and $M$ and $N$ are words of length $n-i$. The sums over deconcatenations of the words $I$ and $J$ into $K$ and $L$ and $M$ and $N$, respectively, as well as the antisymmetrisation between $K$ and $L$, are thus a consequence of expanding these commutators and reordering the terms. 

Using equation \eqref{eqn:3.17} in \eqref{eqn:3.9} gives at last
\begin{align}\label{eq:3.26}
\begin{split}
\Phi &= \sum_{n \geq 1} \sum_{I, J \in \mathcal{W}_n} \phi_{I \vert J} \ue^{\ui k_I \cdot x} T^{\boldsymbol{a}_I} \otimes T'^{\boldsymbol{a}'_J} \\
&= \sum_{i,j \geq 1} \phi_{i \vert j} \ue^{\ui k_{i} \cdot x} T^{a_i} \otimes T'^{a'_j} + \sum_{i,j,k,l \geq 1} \phi_{ij \vert kl} \ue^{\ui k_{ij} \cdot x} T^{a_i} T^{a_j}\otimes T'^{a'_k}T'^{a'_l} + \cdots,
\end{split}
\end{align} 
which, by the foregoing remarks, coincides with the perturbiner expansion for the bi-adjoint scalar theory. It is also worth emphasizing that, by its definition through \eqref{eqn:3.9}, $\Phi$ is a Maurer-Cartan element in $\LBA$. Thus, it satisfies the equation
\begin{equation}
l_1(\Phi) + \frac{1}{2} l_2(\Phi,\Phi) = 0. 
\end{equation}
This, of course, is nothing but the equation of motion \eqref{eqn:2.10}. However, it should be kept in mind that, in contrast with the derivation discussed in \S \ref{sec:2.3}, the recursion relation for the Berends-Giele double currents is encoded in the recursion relations for the $L_{\infty}$-quasi-isomorphism from $H^{\sbullet}(\LBA)$ onto $\LBA$. 

The present formalism also provides a direct means of showing the shuffle constraints for the Berends-Giele double currents $\phi_{I \vert J}$. The starting point is the homotopy Maurer-Cartan equation for $\Phi' \in H^{1}(\LBA) = l_1$,
\begin{equation}\label{eq:3.28}
\sum_{n \geq 2} \frac{1}{n!} l'_{n} (\Phi',\dots,\Phi') = 0. 
\end{equation}
By replacing \eqref{eqn:3.8} in \eqref{eq:3.28}, we see that the latter will be an expansion in the noncommutative variables $\ue^{\ui k_{i}\cdot x} T^{a_i}$ and $T'^{a'_j}$. Thus, equation \eqref{eq:3.28} is solved equating to zero order by order the coefficients in this expansion. To see how this works, it is convenient to introduce, for each positive integer $i$, the formal series
\begin{equation}
\Phi'(i) = \sum_{j \geq 1} \phi_{i \vert j} \ue^{\ui k_{i} \cdot x} T^{a_i} \otimes T'^{a'_j},
\end{equation}
in such a way that $\Phi' = \sum_{i \geq 1} \Phi'(i)$. With this definition, we can rewrite \eqref{eq:3.28} as
\begin{equation}
\sum_{n \geq 2} \sum_{i_1,\dots,i_n \geq 1} \frac{1}{n!}  l'_{n} (\Phi'(i_1),\dots,\Phi'(i_n)) = 0.
\end{equation}
Combining this with equation \eqref{eqn:3.7} then gives
\begin{align}\label{eq:3.31}
\begin{split}
\sum_{n \geq 2} \sum_{i_1,\dots,i_n \geq 1}  \sum_{r=1}^{n-1} \sum_{\sigma \in \mathfrak{S}_{r,n-r}}   \frac{1}{n!}  \big\llbracket f_r (\Phi'(i_{\sigma(1)}) ,\dots, \Phi'(i_{\sigma(r)}) ), f_{n-r}(\Phi'(i_{\sigma(r+1)}) ,\dots, \Phi'(i_{\sigma(n)})) \big\rrbracket  = 0,
\end{split}
\end{align}
where we are abusing notation by identifying elements of $\mathscr{E}(\RR^{1,d-1},\gfrak \otimes \gfrak')$ with their equivalence classes in $\mathscr{E}(\RR^{1,d-1},\gfrak \otimes \gfrak')/\im \left( l_1\right)$  so that $p^{(2)}$ acts trivially. Let us examine the contribution of \eqref{eq:3.31} to the first few orders in $\ue^{\ui k_i \cdot x}T^{a_i}$ and $T'^{a'_j}$. By following the same calculation steps used to derive \eqref{eqn:3.12}, the contribution to second order (modulo constants) takes the form
\begin{align*}
 \big\llbracket  \Phi'(i), \Phi'(j) \big\rrbracket +  \big\llbracket  \Phi'(j), \Phi'(i) \big\rrbracket   &= 2 \sum_{k,l \geq 1} s_{ij}(\phi_{ij \vert kl} + \phi_{ji \vert kl}) \ue^{\ui k_{ij} \cdot x} T^{a_i} T^{a_j} \otimes T'^{a'_k} T'^{a'_l} \\
&= 2  \sum_{k,l \geq 1} s_{ij}\phi_{i \shuffle j \vert kl} \ue^{\ui k_{ij} \cdot x} T^{a_i} T^{a_j} \otimes T'^{a'_k} T'^{a'_l}. 
\end{align*}
Therefore, it must be $\phi_{i \shuffle j \vert kl} = 0$ for all positive integers $i$, $j$, $k$ and $l$. Similarly, the algebraic manipulations that lead to \eqref{eqn:3.14}, yield for the contribution to third order (modulo constants),
\begin{align*}
 \big\llbracket  &\Phi'(i), f_2(\Phi'(j),\Phi'(k)) \big\rrbracket + \big\llbracket  \Phi'(j), f_2(\Phi'(i),\Phi'(k)) \big\rrbracket +  \big\llbracket \Phi'(j), f_2(\Phi'(k),\Phi'(i)) \big\rrbracket \\
&=2 \sum_{l,m,n \geq 1} s_{ijk} (\phi_{ijk \vert lmn} + \phi_{jik \vert lmn} + \phi_{jki \vert lmn}) \ue^{\ui k_{ijk} \cdot x} T^{a_i} T^{a_j} T^{a_k} \otimes T'^{a'_l} T'^{a'_m} T'^{a'_n} \\
&= 2 \sum_{l,m,n \geq 1} s_{ijk} \phi_{i \shuffle jk \vert lmn} \ue^{\ui k_{ijk} \cdot x} T^{a_i} T^{a_j} T^{a_k} \otimes T'^{a'_l} T'^{a'_m} T'^{a'_n}. 
\end{align*}
In consequence, $\phi_{i \shuffle jk \vert lmn} = 0$ for all positive integers $i$, $j$, $k$, $l$, $m$ and $n$. This pattern continues to higher orders. In checking this, it proves very useful to rewrite \eqref{eq:3.31} in the form
\begin{equation}\label{eq:3.32}
\sum_{n \geq 1} \sum_{I \in \mathcal{W}_n} \sum_{r = 1}^{n-1} \sum_{\substack{I = J \shuffle K \\ J \in \mathcal{W}_{r} \\ K \in \mathcal{W}_{n-r}}} \frac{1}{n!} \big\llbracket f_r (\Phi'(J) ), f_{n-r}(\Phi'(K)) \big\rrbracket  = 0. 
\end{equation}
Here, for each pair of words $J = j_1 \cdots j_r$ and $K = k_1 \cdots k_{n-r}$, the symbols $\Phi'(J)$ and $\Phi'(K)$ represent the tuples $(\Phi'(j_{1}), \Phi'(j_{2}), \dots, \Phi'(j_{r}))$ and $(\Phi'(k_{1}), \Phi'(k_{2}), \dots, \Phi'(k_{n-r}))$, respectively. Using equation \eqref{eq:3.32}, it is not difficult to see that the contribution to $n$th order (modulo constants) is none other than
$$
\sum_{L \in \mathcal{W}_n} s_{I} \phi_{J \shuffle K \vert L} \ue^{\ui k_{I} \cdot x} T^{\boldsymbol{a}_I} \otimes T'^{\boldsymbol{a}'_L}.
$$
We have thus come to the conclusion that $\phi_{J \shuffle K \vert L} = 0$ for all words $J$, $K$ and $L$.

%%%%%%%%%%%%%%%%%%%%%%%%%%%%%%%%%%%%%%%%%%%%%%%%%%%%%

\subsection{The Yang-Mills $L_{\infty}$-algebra}\label{sec:3.3}
The Yang-Mills $L_{\infty}$-algebra has been considered in a few papers, e.g., \cite{Movshev:2003ib,Movshev:2004aw,Zeitlin:2007vv,Zeitlin:2007yf}. Let us summarise very briefly its definition following the exposition of \cite{Macrelli:2019afx}. The notation is as in \S \ref{sec:2.4}, except that that we work in dimension $d = 4$. 

Consider the space $\Omega^{r}(\RR^{1,3},\gfrak)$ consisting of $r$-forms on $\RR^{1,3}$ with values in $\gfrak$. We let $\ud$ be the exterior differential, $\ast$ the Hodge star operator induced by the Minkowski metric, and $\delta = \ast \uud \ast$ the corresponding codifferential. Then, the cochain complex underlying the Yang-Mills $L_{\infty}$-algebra $\LYM$ is
$$
\Omega^{0}(\RR^{1,3},\gfrak) \xlongrightarrow{\ud} \Omega^{1}(\RR^{1,3},\gfrak)[-1]  \xlongrightarrow{\delta\ud}  \Omega^{1}(\RR^{1,3},\gfrak)[-2]  \xlongrightarrow{\delta} \Omega^{0}(\RR^{1,3},\gfrak)[-3].
$$
Thus, $\Omega^{0}(\RR^{1,3},\gfrak)$ is situated in degrees $0$ and $3$ and $\Omega^{1}(\RR^{1,3},\gfrak)$ is situated in degrees $1$ and $3$. The non-vanishing higher order brackets are \cite{Macrelli:2019afx}
\begin{equation}\label{eqn:3.18}
\begin{gathered} 
 l_1(c_1) = \ud c_1, \quad l_1(A_1) = \delta \ud A_1, \quad l_1(A_1^{+}) = \delta A_1^{+},  \\
 l_2 (c_1,c_2) = [c_1,c_2], \quad l_2(c_1, A_1) = [c_1,A_1] , \\
 l_2 (c_1,A_2^{+}) = [c_1,A_2^{+}], \quad l_2(c_1, c_2^{+}) = [c_1,c_2^{+}], \\
 l_2 (A_1,A_2^{+}) = \ast [A_1,\ast A_2^{+}] , \\
l_2(A_1,A_2) = \delta [A_1,A_2] + \ast [A_1, \ast d A_2] + \ast [A_2,\ast d A_1], \\
 l_3(A_1,A_2,A_3) = \ast [A_1,\ast [A_2, A_3]] + \ast [A_2,\ast [A_3, A_1]] + \ast [A_3,\ast [A_1, A_2]], 
\end{gathered}
\end{equation}
where $c_1,c_2 \in \Omega^{0}(\RR^{1,3},\gfrak)$, $A_1,A_2,A_3 \in \Omega^{1}(\RR^{1,3},\gfrak)[-1]$, $A_1^{+},A_2^{+} \in \Omega^{1}(\RR^{1,3},\gfrak)[-2]$ and $c_2^{+} \in \Omega^{0}(\RR^{1,3},\gfrak)[-3]$ and $[,]$ denotes the graded Lie bracket on $\Omega^{r}(\RR^{1,3},\gfrak)$ given by combining the wedge product and the Lie bracket on $\gfrak$. This $L_{\infty}$-algebra admits a cyclic inner product that is non-vanishing only when the total degree is $3$, and is consequently determined by
\begin{align} \label{eqn:3.19}
\begin{split}
\langle c, c^{+} \rangle &= - \int_{\RR^{1,3}} \ud^4x \, c^{a} c^{+}_{a}, \\
\langle A, A^{+} \rangle &= - \int_{\RR^{1,3}} \ud^4 x \, A^{a}_{\mu} A^{+\mu}_a,
\end{split}
\end{align}
where $c \in \Omega^{0}(\RR^{1,3},\gfrak)$, $c^{+} \in \Omega^{0}(\RR^{1,3},\gfrak)[-3]$, $A \in \Omega^{1}(\RR^{1,3},\gfrak)[-1]$ and $A^{+} \in \Omega^{1}(\RR^{1,3},\gfrak)[-2]$, and where, as usual, $c = c_{a} T^{a}$, $c^{+} = c^{+}_a T^{a}$ and similarly for $A_{\mu}$ and $A^{+}_{\mu}$. With this definition, it is not hard to see that, for all $A \in \Omega^{1}(\RR^{1,3},\gfrak)[-1]$,
\begin{align*}
\frac{1}{2} \langle A, l_1(A) \rangle &= - \frac{1}{4} \int_{\RR^{1,3}} \ud^4 x \, (\partial_{\mu} A^{a}_{\nu} - \partial_{\nu} A^{a}_{\mu})(\partial^{\mu} A^{\nu}_a - \partial^{\nu} A^{\mu}_a), \\
\frac{1}{3!} \langle A, l_2(A,A) \rangle &= - \frac{1}{2} \int_{\RR^{1,3}} \ud^4 x \, f^{bc}_{\phantom{bc}a}(\partial_{\mu} A^{a}_{\nu} - \partial_{\nu} A^{a}_{\mu}) A^{\mu}_b A^{\nu}_c, \\
\frac{1}{4!} \langle A, l_3(A,A,A) \rangle &= - \frac{1}{4} \int_{\RR^{1,3}} \ud^4 x \, f^{bc}_{\phantom{bc}a} f^{dea} A_{\mu b} A_{\nu c} A^{\mu}_d A^{\nu}_e,
\end{align*}
from which the homotopy Maurer-Cartan action for the $L_{\infty}$-algebra $\LYM$ is deduced to be
$$
S_{\mathrm{MC}}[A] = - \frac{1}{4}  \int_{\RR^{1,3}} \ud^4 x \, F^{a}_{\mu\nu} F^{\mu\nu}_a.
$$
This is simply the familiar Yang-Mills action \eqref{eqn:2.13}.

Now, just as in the case of the bi-adjoint scalar theory, to deal with the perturbiner expansion we need to modify a little bit the definition of the $L_{\infty}$-algebra $\LYM$. Let us keep fixed from now an infinite multiset of ``colour indices'' $\boldsymbol{a} = (a_{i})_{i \geq 1}$ and an infinite set $(k_{i})_{i \geq 1}$ of massless momentum vectors in $\RR^{1,3}$. We denote by $\mathscr{E}^{0}(\RR^{1,3},\gfrak)$ the space of formal series of the form
\begin{align}\label{eqn:3.20}
\begin{split}
h(x) = \sum_{m \geq 1} \sum_{I \in \mathcal{W}_m} h_{I} \ue^{\ui k_{I} \cdot x} T^{\boldsymbol{a}_I} = \sum_{i \geq 1} h_{i} \ue^{\ui k_{i} \cdot x} T^{a_i} + \sum_{i, j \geq 1} h_{ij} \ue^{\ui k_{ij} \cdot x} T^{a_i} T^{a_j} + \cdots,
\end{split}
\end{align}
and by $\mathscr{E}^{r}(\RR^{1,3},\gfrak)$ the space of $r$-forms on $\RR^{1,3}$ with coefficients on $\mathscr{E}^{0}(\RR^{1,3},\gfrak)$. In keeping with the terminology suggested by \cite{Mizera:2018jbh}, elements of $\mathscr{E}^{1}(\RR^{1,3},\gfrak)$ may be called \emph{colour-stripped perturbiner ansatzs}. 

We would next like to extend the exterior differential $\ud$, the Hodge star operator $\ast$ and the codifferential $\delta$ to all of $\mathscr{E}^{\sbullet}(\RR^{1,3},\gfrak)$. To do this, we can again resort to the colour-dressed version of the elements of each space $\mathscr{E}^{r}(\RR^{1,3},\gfrak)$. In the standard coordinates of $\RR^{1,3}$, an element of $\mathscr{E}^{r}(\RR^{1,3},\gfrak)$ is written $C(x) = \frac{1}{r!} C_{\mu_1 \cdots \mu_r}(x) \ud x^{\mu_1} \wedge \cdots \wedge \ud x^{\mu_r}$ where, in accord with \eqref{eqn:3.20}, the components $C_{\mu_1 \cdots \mu_r}(x)$, or rather the associated components $C^{\mu_1 \cdots \mu_r}(x)$, are formal series of the form
\begin{align}\label{eqn:3.21}
\begin{split}
C^{\mu_1 \cdots \mu_r}(x) &= \sum_{m \geq 1} \sum_{I \in \mathcal{W}_m} C^{\mu_1 \cdots \mu_r }_I \ue^{\ui k_{I} \cdot x} T^{\boldsymbol{a}_I} \\
&= \sum_{i \geq 1} C^{\mu_1 \cdots \mu_r}_{i} \ue^{\ui k_{i} \cdot x} T^{a_i} + \sum_{i, j \geq 1} C^{\mu_1 \cdots \mu_r}_{ij} \ue^{\ui k_{ij} \cdot x} T^{a_i} T^{a_j} + \cdots.
\end{split}
\end{align} 
With this in mind, for each sequence of positive integers $i_1 < i_2 < \cdots < i_m$, we set
\begin{equation}
f^{i_1 i_2 \cdots i_m}_{\phantom{i_1 i_2 \cdots i_m} a} = f^{a_{i_1} a_{i_2}}_{\phantom{a_{i_1} a_{i_2}} b} f^{b a_{i_3}}_{\phantom{b a_{i_3}} c} \cdots f^{d a_{i_{m-1}}}_{\phantom{d a_{i_{m-1}}} e} f^{e a_{i_{m}}}_{\phantom{e a_{i_{m}}} a},
\end{equation}
and then define
\begin{equation}
C^{\mu_1 \cdots \mu_r}_{a \vert i_1 i_2 \cdots i_m} = \sum_{\sigma \in \mathfrak{S}_{m-1}} f^{i_1 i_{\sigma(2)} \cdots i_{\sigma(m)}}_{\phantom{i_1 i_{\sigma(2)} \cdots i_{\sigma(m)}} a} C^{\mu_1 \cdots \mu_r}_{i_1 i_{\sigma(2)} \cdots i_{\sigma(m)}}. 
\end{equation}
Using the latter, equation \eqref{eqn:3.21} can be rewritten as $C^{\mu_1 \cdots \mu_r}(x) = C^{\mu_1 \cdots \mu_r}_{a}(x) T^{a}$, where the the coefficients $C^{\mu_1 \cdots \mu_r}_{a}(x)$ are formal series of the form
\begin{align}
\begin{split}
C^{\mu_1 \cdots \mu_r}_{a}(x) &= \sum_{m \geq 1} \sum_{I \in \mathcal{OW}_m} C^{\mu_1 \cdots \mu_r}_{a \vert I} \ue^{\ui k_I \cdot x} \\
&= \sum_{i} C^{\mu_1 \cdots \mu_r}_{a \vert i} \ue^{\ui k_i \cdot x} + \sum_{i < j} C^{\mu_1 \cdots \mu_r}_{a \vert ij} \ue^{\ui k_{ij} \cdot x} + \sum_{i < j < k} C^{\mu_1 \cdots \mu_r}_{a \vert ijk} \ue^{\ui k_{ijk} \cdot x} + \cdots. 
\end{split}
\end{align}
With this expression in hand, it is now straightforward to extend the definition of the exterior differential $\ud$, the Hodge star operator $\ast$ and the codifferential $\delta$ to the spaces $\mathscr{E}^{\sbullet}(\RR^{1,3},\gfrak)$. 

From the above discussion, the cyclic $L_{\infty}$-algebra $\LYM$ controlling the perturbiner expansion for the Yang-Mills theory is described by the cochain complex 
\begin{equation*}
\mathscr{E}^{0}(\RR^{1,3},\gfrak) \xlongrightarrow{\ud} \mathscr{E}^{1}(\RR^{1,3},\gfrak)[-1]  \xlongrightarrow{\delta \ud}  \mathscr{E}^{1}(\RR^{1,3},\gfrak)[-2]  \xlongrightarrow{\delta} \mathscr{E}^{0}(\RR^{1,3},\gfrak)[-3].
\end{equation*}
The higher order brackets and the cyclic inner product are determined by the same formulas as in \eqref{eqn:3.18} and \eqref{eqn:3.19}. 

%%%%%%%%%%%%%%%%%%%%%%%%%%%%%%%%%%%%%%%%%%%%%%%%%%%%%%%%%%%%

\subsection{The perturbiner expansion for Yang-Mills theory revisited}\label{sec:3.4}
We shall now proceed to show how the perturbiner expansion for the Yang-Mills theory is determined by the minimal model for the Yang-Mills $L_{\infty}$-algebra $\LYM$. The approach and calculations are very similar to the ones used in \S \ref{sec:3.2} and \cite{Macrelli:2019afx}, so some details are skipped. 

First of all, we must notice that, just as in the ordinary case, there is an abstract Hodge-Kodaira decomposition on the cohomology $H^{\sbullet}(\LYM)$ of $\LYM$ (see Appendix B of \cite{Jurco:2018sby}). We may then use this decomposition to show that the cochain complex underlying $H^{\sbullet}(\LYM)$ is
$$
 \ker \left( \ud \right) \xlongrightarrow{0} \ker \left( \delta \ud\right) / \im \left( \ud \right)[-1] \xlongrightarrow{0} \ker \left( \delta \ud\right) / \im \left( \ud \right)[-2] \xlongrightarrow{0} \ker \left(  \ud\right) [-3] . 
$$
The projections $p^{(0)}, p^{(3)} \colon \mathscr{E}^{0}(\RR^{1,3},\gfrak) \to  \ker \left( \ud \right)$ and $p^{(1)}, p^{(2)} \colon \mathscr{E}^{1}(\RR^{1,3},\gfrak) \to  \ker \left( \delta \ud\right) / \im \left( \ud \right)$ are thus chosen to be the natural projections induced by the Hodge-Kodaira decomposition, and similarly, the embeddings $i^{(0)}, i^{(3)} \colon  \ker \left( \ud \right) \to \mathscr{E}^{0}(\RR^{1,3},\gfrak)$ and $i^{(1)}, i^{(2)} \colon  \ker \left( \delta \ud\right) / \im \left( \ud \right) \to \mathscr{E}^{1}(\RR^{1,3},\gfrak)$ are chosen to be the trivial ones. On the other hand, to define a contracting homotopy $h \colon \LYM \to \LYM$, we must extend the Feynman propagator $\GF$ considered in \S \ref{sec:3.2} in such a way to get a linear operator $\GF \colon \mathscr{E}^r(\RR^{1,3},\gfrak) \to \mathscr{E}^r(\RR^{1,3},\gfrak)$. We also need to consider the projector $P_{\ue} \colon \mathscr{E}(\RR^{1,3},\gfrak) \to  \mathscr{E}(\RR^{1,3},\gfrak)$ onto the image of $\delta \ud$. In terms of these, the three non-zero components of the contracting homotopy $h$ read as
\begin{equation*}
\begin{alignedat}{2}
h^{(1)} &=  \GF \circ \delta &\colon \mathscr{E}^1(\RR^{1,3},\gfrak) \lto \mathscr{E}^0(\RR^{1,3},\gfrak), \\
h^{(2)} &=  \GF \circ P_{\ue} &\colon \mathscr{E}^1(\RR^{1,3},\gfrak) \lto \mathscr{E}^1(\RR^{1,3},\gfrak), \\
h^{(3)} &= \GF \circ \uud &\colon \mathscr{E}^0(\RR^{1,3},\gfrak) \lto \mathscr{E}^1(\RR^{1,3},\gfrak).
\end{alignedat}
\end{equation*}
It should be mentioned that the formulas for the $L_{\infty}$-quasi-isomorphism and the higher-order brackets for the minimal $L_{\infty}$-structure on $H^{\sbullet}(\LYM)$ are derived under the assumption that $h^{(1)}(A)= 0$ for any $A \in \mathscr{E}^1(\RR^{1,3},\gfrak)$ (see Appendix A of \cite{Macrelli:2019afx}). This implies that the colour-stripped perturbiner ansatzs satisfy the Lorenz gauge condition $\delta A = 0$. Moreover, the second component of the contracting homotopy $h^{(2)} = \GF \circ P_{\ue}$ must be interpreted as the gluon propagator. 

It is now easy to adapt the formulas given in  \cite[\S 2.1]{Macrelli:2019afx} for the $L_{\infty}$-quasi-isomorphism between $\LYM$ and $H^{\sbullet}(\LYM)$. These are just generalisations of the formulas exhibited in \eqref{eqn:2.5} that include the higher order bracket $l_3$. For the sake of clarity, we shall only write them down in homogeneous degree $1$. Therefore, let us fix $A'_1,\dots,A'_n \in H^{1}(\LYM) = \ker \left( \delta \ud \right) / \im \left( \ud \right)$. Then the action of $f_n$ on $A'_1,\dots, A'_n$ is characterised by the formula
\begin{align}\label{eqn:3.23}
\begin{split}
f_n (&A'_1,\dots, A'_n) \\  & = - \tfrac{1}{2} \sum_{i+j =n} \sum_{\sigma \in \mathfrak{S}_{i,j}} (h^{(2)} \circ l_2) \left(f_{i}(A'_{\sigma(1)}, \dots, A'_{\sigma(i)}), f_{j}(A'_{\sigma(i+1)}, \dots, A'_{\sigma(n)} \right) \\
& \quad  -\tfrac{1}{3!} \sum_{i+j+k =n} \sum_{\sigma \in \mathfrak{S}_{i,j,k}} (h^{(2)} \circ l_3)\left(f_{i}(A'_{\sigma(1)}, \dots, A'_{\sigma(i)}), f_{j}(A'_{\sigma(i+1)}, \dots, A'_{\sigma(i+j)}), \right. \\
&\quad \qquad\qquad\qquad\qquad\qquad\qquad\quad\, \left. f_{k}(A'_{\sigma(i+j +1)}, \dots, A'_{\sigma(n)}) \right). 
\end{split}
\end{align} 
Similar remarks could be made for the higher order brackets on $H^{\sbullet}(\LYM)$. Again, for the sake of clarity, we shall only display such bracket among the $A'_1,\dots, A'_n$. It is given by
\begin{align}\label{eqn:3.24}
\begin{split}
l'_n (&A'_1,\dots, A'_n) \\  & =  \tfrac{1}{2} \sum_{i+j =n} \sum_{\sigma \in \mathfrak{S}_{i,j}} (p^{(2)} \circ l_2) \left(f_{i}(A'_{\sigma(1)}, \dots, A'_{\sigma(i)}), f_{j}(A'_{\sigma(i+1)}, \dots, A'_{\sigma(n)} \right) \\
& \quad  + \tfrac{1}{3!} \sum_{i+j+k =n} \sum_{\sigma \in \mathfrak{S}_{i,j,k}} (p^{(2)} \circ l_3)\left(f_{i}(A'_{\sigma(1)}, \dots, A'_{\sigma(i)}), f_{j}(A'_{\sigma(i+1)}, \dots, A'_{\sigma(i+j)}), \right. \\
&\quad \qquad\qquad\qquad\qquad\qquad\qquad\quad\, \left. f_{k}(A'_{\sigma(i+j +1)}, \dots, A'_{\sigma(n)}) \right). 
\end{split}
\end{align}
All this must be supplemented with the constraints imposed by \eqref{eqn:2.2} and \eqref{eqn:2.3}. 

We are now ready to come properly to our main objective herewith, that is, to examine how the perturbiner expansion for the Yang-Mills theory can be extracted from the minimal $L_{\infty}$-structure on $H^{\sbullet}(\LYM)$. In analogy with the bi-adjoint scalar theory case, we start with a Maurer-Cartan element $A' \in H^{1}(\LYM) = \ker \left( \delta \ud \right) / \im \left( \ud \right)$ with components with respect to the standard coordinates of $\RR^{1,3}$ in the form of a plane-wave superposition of all particles:
\begin{equation}\label{eq:3.42}
A'^{\mu} = \sum_{i \geq 1} \mathcal{A}^{\mu}_{i} \ue^{\ui k_i \cdot x} T^{a_i}. 
\end{equation}
Then we claim that the perturbiner expansion is determined by the element $A \in \mathscr{E}(\RR^{1,3},\gfrak)[-1]$ with components
\begin{equation} \label{eqn:3.26}
A^{\mu} = \sum_{n \geq 1} \frac{1}{n!} f_{n}(A',\dots,A')^{\mu}. 
\end{equation}
As before, to verify this, we just need to calculate explicitly each term of this sum. 

To this end, we first notice that the inner sums in \eqref{eqn:3.23} have, respectively, $\binom{n}{i}$ and $\binom{n}{i}\binom{n-i}{j}$ summands, and as a consequence
\begin{align}\label{eqn:3.27}
\begin{split}
f_n (A',\dots, A') & = - \tfrac{1}{2} \sum_{i+j =n} \binom{n}{i} (h^{(2)} \circ l_2) \left(f_{i}(A', \dots, A'), f_{j}(A', \dots, A') \right) \\
& \quad  -\tfrac{1}{3!} \sum_{i+j+k =n} \binom{n}{i}\binom{n-i}{j} (h^{(2)} \circ l_3)\left(f_{i}(A', \dots, A'), f_{j}(A', \dots, A'), \right. \\
&\qquad \qquad\qquad\qquad\qquad\qquad\qquad\quad\qquad \left.  f_{k}(A', \dots, A') \right). 
\end{split}
\end{align}
For $n = 2$ this gives
$$
f_2 (A',A') = -\tfrac{1}{2} \binom{2}{1} (\GF \circ P_{\ue}) \left( l_2(f_1(A'),f_1(A')) \right) = - \GF \left( l_2(A', A') \right),
$$
where in the second equality, we identify the element $A'$ with its image $f_1(A')$ and used that $P_{\ue}$ acts trivially. At the same time, attending to the definition in \eqref{eqn:3.18}, we get
$$
l_2 (A',A')^{\mu} = 2 \ui \left\{ \partial_{\nu} [A'^{\nu},A'^{\mu}] + [A'^{\nu}, \partial_{\nu} A'^{\mu} - \partial^{\mu} A'_{\nu}] \right\}.  
$$
Therefore, by a direct computation, the components of $f_2(A',A')$ result in the form
$$
f_2(A',A')^{\mu} = 2! \sum_{i,j \geq 1} \mathcal{A}_{ij}^{\mu} \ue^{\ui k_{ij} \cdot x} T^{a_i} T^{a_j},
$$
with coefficients determined by 
$$
\mathcal{A}^{\mu}_{ij} = \frac{1}{s_{ij}} \big\{ (\mathcal{A}_i \cdot k_i) \mathcal{A}^{\mu}_j + \mathcal{A}_{i \nu} \mathcal{F}^{\mu \nu}_{j} - (\mathcal{A}_j \cdot k_j) \mathcal{A}^{\mu}_i - \mathcal{A}_{j \nu} \mathcal{F}^{\mu \nu}_{i}  \big\},
$$
where we have set $\mathcal{F}^{\mu \nu}_{j} = k^{\mu}_i \mathcal{A}^{\nu}_i - k^{\nu}_i \mathcal{A}^{\mu}_i$. For $n = 3$, equation \eqref{eqn:3.27} is already more complicated, and reads
\begin{align*}
f_3 (A',A',A') &= -\tfrac{1}{2} \binom{3}{1} (\GF \circ P_{\ue}) \left( l_2 (f_1(A'), f_{2}(A',A'))\right) \\
& \quad -\tfrac{1}{2} \binom{3}{2} (\GF \circ P_{\ue}) \left( l_2 (f_{2}(A',A'),f_1(A'))\right) \\
&\quad -\tfrac{1}{3!} \binom{3}{1}\binom{2}{1} (\GF \circ P_{\ue}) \left( l_3(f_1(A'),f_1(A'),f_1(A')) \right) \\
&= - 3 \GF \left(  l_2 (f_{1}(A'), f_{2}(A',A'))\right) - \GF \left( l_3(f_1(A'),f_1(A'),f_1(A')) \right).
\end{align*}
Upon using $l_3 (A',A',A')^{\mu} = - 3! [A'_{\nu}, [A'^{\nu}, A'^{\mu}]]$, which again follows from the definition in \eqref{eqn:3.18}, we find, after a few calculations, that the components of $f_3(A',A',A')$ are
$$
f_3(A',A',A') = 3! \sum_{i,j,k \geq 1} \mathcal{A}^{\mu}_{ijk} \ue^{k_{ijk} \cdot x} T^{a_i}T^{a_j}T^{a_k},
$$
with coefficients given by
\begin{align*}
\mathcal{A}^{\mu}_{ijk} = \frac{1}{s_{ijk}} \big\{ &(k_{i} \cdot \mathcal{A}_{i} ) \mathcal{A}^{\mu}_{jk} + \mathcal{A}_{i\nu} \mathcal{F}^{\mu\nu}_{jk} - (k_{jk} \cdot \mathcal{A}_{jk} ) \mathcal{A}^{\mu}_{i} - \mathcal{A}_{jk\nu} \mathcal{F}^{\mu\nu}_{i} \\
& + (k_{ij} \cdot \mathcal{A}_{ij} ) \mathcal{A}^{\mu}_{k} + \mathcal{A}_{ij\nu} \mathcal{F}^{\mu\nu}_{k} - (k_{k} \cdot \mathcal{A}_{k} ) \mathcal{A}^{\mu}_{ij} - \mathcal{A}_{k\nu} \mathcal{F}^{\mu\nu}_{ij} \big\},
\end{align*}
where we have put $\mathcal{F}^{\mu\nu}_{ij} = k^{\mu}_{ij} \mathcal{A}^{\nu}_{ij} - k^{\nu}_{ij} \mathcal{A}^{\mu}_{ij} - \mathcal{A}^{\mu}_{i} \mathcal{A}^{\nu}_{j} + \mathcal{A}^{\mu}_{j} \mathcal{A}^{\nu}_{i}$. This pattern continues as we keep increasing the value of $n$. Indeed, using mathematical induction we can prove that the components of $f_n (A',\dots,A')$ are
\begin{equation}\label{eqn:3.28}
f_{n}(A',\dots,A')^{\mu} = n! \sum_{I \in \mathcal{W}_n} \mathcal{A}^{\mu}_I \ue^{\ui k_I \cdot x} T^{\boldsymbol{a}_I},
\end{equation} 
where the coefficients $\mathcal{A}^{\mu}_I$ are determined from the recursion relation \eqref{eqn:2.16} for the Berend-Giele currents. Inserting \eqref{eqn:3.28} back in \eqref{eqn:3.26}, we finally get
\begin{align}
\begin{split}
A^{\mu} &= \sum_{n \geq 1} \sum_{I \in \mathcal{W}_n} \mathcal{A}^{\mu}_I \ue^{\ui k_{I} \cdot x} T^{\boldsymbol{a}_I} = \sum_{i \geq 1} \mathcal{A}^{\mu}_i \ue^{\ui k_{i} \cdot x} T^{a_i} + \sum_{i,j \geq 1} \mathcal{A}^{\mu}_{ij} \ue^{\ui k_{ij} \cdot x} T^{a_{i}}T^{a_{j}} + \cdots ,
\end{split}
\end{align}
which agrees with the perturbiner expansion for the Yang-Mills theory, as was to be shown. It should also be borne in mind that the element $A$ in $\LYM$ with components defined by \eqref{eqn:3.26} is Maurer-Cartan, and therefore satisfies the equation
\begin{equation}
l_1(A)^{\mu} + \frac{1}{2} l_2(A,A)^{\mu} + \frac{1}{3!} l_3(A,A,A)^{\mu} = 0. 
\end{equation}
Since $l_1(A)^{\mu} = \square A^{\mu}$, it is seen that the latter coincides precisely with the equation of motion \eqref{eqn:2.14}. Note, however, that as in the bi-adjoint scalar theory case and in contrast to the discussion in \S \ref{sec:2.4}, the recursion relation for the Berends-Giele currents is here encoded in the recursion relation for the $L_{\infty}$-quasi-isomorphism between $H^{\sbullet}(\LYM)$ and $\LYM$. 

To close this subsection, we wish to point out that, just as in our earlier treatment in \S \ref{sec:3.2}, the shuffle constraints for the Berends-Giele currents $\mathcal{A}_{I}^{\mu}$ follow from the homotopy Maurer-Cartan equation for $A' \in H^{1}(\LYM) = \ker \left( \delta \ud \right) / \im \left( \ud \right)$ corresponding to the higher order brackets described in \eqref{eqn:3.24}. We leave it to the reader to fill in the details.

\section{Scattering amplitudes}\label{sec:4}
As explained in \cite{Mafra:2015vca}, there is a generating series of tree-level gluon amplitudes in super Yang-Mills theory, which matches the ten-dimensional lagrangian evaluated on a perturbiner-like generating series of Berends-Giele currents in superspace. In accordance with this result, in this section we will show how the perturbiner expansions for the bi-adjoint scalar and Yang-Mills theory play a more active role by effectively generating the tree-level scattering amplitudes of the theory directly from the homtopy Maurer-Cartan action.  
 
\subsection{Bi-adjoint scalar theory}
Let us start by considering the tree-level scattering amplitudes for the bi-adjoint scalar theory. The notation is the same as in \S \ref{sec:3.1} and \S \ref{sec:3.2}.

The basic idea is to insert the multi-particle solution \eqref{eqn:3.8} into the homotopy Maurer-Cartan action
\begin{equation}\label{eq:4.1}
S'_{\mathrm{MC}}[\Phi'] = \sum_{n \geq 2} \frac{1}{(n+1)!} \langle \Phi' , l'_n (\Phi',\dots,\Phi') \rangle. 
\end{equation}
To determine the sum, we use \eqref{eqn:3.7}, which gives the higher order bracket as
\begin{equation}\label{eq:4.2}
l'_n (\Phi',\dots,\Phi') = \tfrac{1}{2} \sum_{i =1}^{n-1} \binom{n}{i} \big\llbracket f_i (\Phi',\dots,\Phi'), f_{n-i}(\Phi',\dots,\Phi') \big\rrbracket.
\end{equation}
Here, we are again `abusing notation' by suppressing the action of $p^{(2)}$. By following the same arguments as those employed in deriving \eqref{eqn:3.17}, equation \eqref{eq:4.2} becomes
\begin{equation}\label{eq:4.3}
l'_n (\Phi',\dots,\Phi') = -n! \sum_{I,J \in \mathcal{W}_n} s_I \phi_{I \vert J} \ue^{\ui k_{I} \cdot x} T^{\boldsymbol{a}_I} \otimes T'^{\boldsymbol{a}'_J}. 
\end{equation}
Thus, on account of \eqref{eqn:3.2}, we have
\begin{align*}
 \langle \Phi' , l'_n (\Phi',\dots,\Phi') \rangle &= -n! \sum_{i,j \geq 1} \sum_{I,J \in \mathcal{W}_n} \int_{\RR^{1,d-1}} \ud^d x \, \phi_{i \vert j} \ue^{\ui k_{i} \cdot x} s_{I} \phi_{I\vert J} \ue^{\ui k_{I} \cdot x} \tr (T^{a_i} T^{\boldsymbol{a}_I}) \tr (T^{a'_j} T^{\boldsymbol{a}'_J}) \\
 &=  -n! \sum_{i,j \geq 1} \sum_{I,J \in \mathcal{W}_n} \int_{\RR^{1,d-1}} \ud^d x \, s_I \phi_{i \vert j} \phi_{I\vert J} \ue^{\ui k_{iI} \cdot x}\tr (T^{\boldsymbol{a}_{iI}}) \tr (T^{\boldsymbol{a}'_{jJ}})  \\
 &=-(2 \pi)^d n!  \sum_{i,j \geq 1} \sum_{I,J \in \mathcal{W}_n} \delta(k_{iI}) s_I \phi_{i \vert j} \phi_{I\vert J} \tr (T^{\boldsymbol{a}_{iI}}) \tr (T^{\boldsymbol{a}'_{jJ}}),
\end{align*}
which, by virtue of \eqref{eq:2.27} and \eqref{eq:2.28}, translates to
\begin{equation}\label{eq:4.4}
 \langle \Phi' , l'_n (\Phi',\dots,\Phi') \rangle = -(2 \pi)^d (n+1)! \mathscr{M}_{n+1}^{\mathrm{tree}}.
\end{equation}
Substituting this back into \eqref{eq:4.1} then gives
\begin{equation}\label{eq:4.5}
S'_{\mathrm{MC}}[\Phi'] = -(2 \pi)^d \sum_{n \geq 3}  \mathscr{M}_n^{\mathrm{tree}}.
\end{equation}
We therefore conclude that the $n$-point tree-level scattering amplitudes $\mathscr{M}_n^{\mathrm{tree}}$ follow directly from the homotopy Maurer-Cartan action for the minimal $L_{\infty}$-structure on $H^{\sbullet}(\LBA)$, evaluated at the plane-wave superposition $\Phi'$. 

Now, what about the perturbiner expansion $\Phi$? If we plug it in the homotopy Maurer-Cartan action for the bi-adjoint DG Lie algebra $\LBA$, we get
\begin{equation}\label{eq:4.6}
S_{\mathrm{MC}}[\Phi] = \frac{1}{2} \langle \Phi, l_1(\Phi) \rangle + \frac{1}{3!} \langle \Phi, l_2(\Phi,\Phi) \rangle. 
\end{equation}
By using the relation \eqref{eqn:3.9}, we can show through a direct but tedious calculation that the pull-back of the homotopy Maurer-Cartan action \eqref{eq:4.6} by the $L_{\infty}$-quasi-isomorphism $f \colon H^{\sbullet}(\LBA) \to \LBA$ coincides with the homotopy Maurer-Cartan action \eqref{eq:4.1}; see, for instance, \S 5.5 of \cite{Kajiura:2003ax}. In symbols,
\begin{equation}\label{eq:4.7}
S'_{\mathrm{MC}}[\Phi'] = f^{*} S_{\mathrm{MC}} [\Phi]. 
\end{equation}
Also, from \emph{loc.~cit.}, one could see that
\begin{equation}\label{eq:4.8}
f^* S_{\mathrm{MC}}[\Phi] = (S_{\mathrm{MC}}[\Phi])_{\mathrm{c}},
\end{equation}
where the subscript `$\mathrm{c}$' denotes the operation of cyclic symmetrisation. Combining \eqref{eq:4.5} with \eqref{eq:4.7} and \eqref{eq:4.8}, we arrive at the result that, up to cyclic ordering of the multi-indices labelling the doubly colour-ordered partial amplitudes, 
\begin{equation}
S_{\mathrm{MC}}[\Phi] \simeq -(2 \pi)^d \sum_{n \geq 3}  \mathscr{M}_n^{\mathrm{tree}}.
\end{equation}
This is a non-trivial result because it indicates that the action for the bi-adjoint scalar theory may be expanded as a sum of the tree-level scattering amplitudes by simply inserting in it the perturbiner expansion.

\subsection{Yang-Mills theory}
Now let us turn to the tree-level scattering amplitudes for Yang-Mills theory. The notation of \S \ref{sec:3.3} and \S \ref{sec:3.4} will be left intact. 

As in the preceding subsection, the starting point is to insert the plane-wave superposition \eqref{eq:3.42} into the homotopy Maurer-Cartan action
\begin{equation}\label{eq:4.10}
S'_{\mathrm{MC}}[A'] = \sum_{n \geq 2} \frac{1}{(n+1)!} \langle A', l'_n (A',\dots,A') \rangle. 
\end{equation}
Next, we must use equation \eqref{eqn:3.24} to evaluate the higher order brackets $l'_n (A',\dots,A')$. By precisely the same argument used to derive \eqref{eqn:3.28}, we find that
\begin{equation}
l'_n (A',\dots,A')^{\mu} = -n! \sum_{I \in \mathcal{W}_n} s_{I} \mathcal{A}_I^{\mu} \ue^{\ui k_{I} \cdot x} T^{\boldsymbol{a}_I}.
\end{equation}
Hence, by use of 	\eqref{eqn:3.19},
\begin{align*}
\langle A' , l'_n (A',\dots,A') \rangle &= n! \sum_{i \geq 1} \sum_{I \in \mathcal{W}_n} \int_{\RR^{1,3}} \ud^4 x \, \mathcal{A}_{i \mu} \ue^{\ui k_i \cdot x} s_I \mathcal{A}_{I}^{\mu} \ue^{\ui k_I \cdot x} \tr (T^{a_i} T^{\boldsymbol{a}_I}) \\
&= n! \sum_{i \geq 1} \sum_{I \in \mathcal{W}_n} \int_{\RR^{1,3}} \ud^4 x \, s_I  \mathcal{A}_i \cdot  \mathcal{A}_I \ue^{\ui k_{iI} \cdot x} \tr(T^{\boldsymbol{a}_{iI}}) \\
&= (2 \pi)^{4} n! \sum_{i \geq 1} \sum_{I \in \mathcal{W}_n} \delta(k_{iI}) s_I  \mathcal{A}_i \cdot  \mathcal{A}_I \tr(T^{\boldsymbol{a}_{iI}}),
\end{align*}
and with \eqref{eq:2.35} and \eqref{eq:2.36}, this tells us that
\begin{equation}
\langle A' , l'_n (A',\dots,A') \rangle = (2 \pi)^4 (n+1)! \mathscr{A}_{n+1}^{\mathrm{tree}}. 
\end{equation}
Now substitute back this expression into \eqref{eq:4.10} to obtain
\begin{equation}\label{eq:4.13}
S'_{\mathrm{MC}}[A'] = (2 \pi)^4 \sum_{n \geq 3} \mathscr{A}_n^{\mathrm{tree}},
\end{equation}
where we have used the fact that $\mathscr{A}_2^{\mathrm{tree}} = 0$. Consequently, just as in the bi-adjoint scalar case, the $n$-point tree-level scattering amplitudes $\mathscr{A}_n^{\mathrm{tree}}$ are obtained by evaluating at the plane-wave superposition $A'$ the homotopy Maurer-Cartan action for the minimal $L_{\infty}$-structure on $H^{\sbullet}(\LYM)$. 

In order to bring to bear the perturbiner expansion $A$, we insert it into the homotopy Maurer-Cartan action for the Yang-Mills $L_{\infty}$-algebra $\LYM$ to yield
\begin{equation}
S_{\mathrm{MC}}[A] = \frac{1}{2} \langle A, l_1(A) \rangle + \frac{1}{3!} \langle A, l_2(A,A) \rangle + \frac{1}{4!} \langle A, l_3(A,A,A) \rangle. 
\end{equation}
It follows then by precisely the same argument used in the previous subsection that
\begin{equation}\label{eq:4.15}
S'_{\mathrm{MC}}[A'] = (S_{\mathrm{MC}}[A])_{\mathrm{c}}.  
\end{equation}
Putting together equations \eqref{eq:4.13} and \eqref{eq:4.15}, we see that, up to cyclic ordering of the multi-indices labelling the colour-ordered partial amplitudes,
\begin{equation}
S_{\mathrm{MC}}[A] \simeq (2 \pi)^4 \sum_{n \geq 3}  \mathscr{A}_n^{\mathrm{tree}}. 
\end{equation}
We therefore conclude that the Yang-Mills action can be expanded in terms of the three-level scattering amplitudes provided that we insert in it the perturbiner expansion. 

%%%%%%%%%%%%%%%%%%%%%%%%%%%%%%%%%%%%%%%%%%

\section{Conclusion and outlook}\label{sec:5}
In this article, we have shown that the perturbiner expansions for the bi-adjoint scalar and Yang-Mills theories can be extracted from the $L_{\infty}$-quasi-isomorphisms to the minimal models of their corresponding $L_{\infty}$-algebras. We have also shown that the tree-level scattering amplitudes can be obtained by substituting such perturbiner expansions into the associated homtopy Maurer-Cartan actions. This renders an alternative new interpretation of the perturbiner formalism, which is in contrast to the usual treatment, where the perturbiner expansion is set as an ansatz for a solution of the non-linear equations of motion that leads to the Berends-Giele recursion relations. 

Now that we have a different take on perturbiner expansions and tree-level scattering amplitudes from an $L_{\infty}$-algebra perspective, there are a number of interesting avenues to pursue. To begin with, it will be worth extending the method to the theory of pure gravity, theories coupled to gravitational backgrounds and Yang-Mills theories in interaction with matter fields. Further, it would be most interesting to examine the case of super Yang-Mills theory in the BCJ gauge along the lines of \cite{Bridges:2019siz}. Going forward, it is also our aim to eventually have an inside view into the double copy relation between gravity and Yang-Mills theories. Finally, another aspect that should be explored is the possible connection between the $L_{\infty}$-algebra technology and  the geometric approaches to computing scattering amplitudes based on the associahedron and the amplituhedron \cite{Arkani-Hamed:2013jha,Arkani-Hamed:2017mur}.

\subsection*{Acknowledgements}
The authors would like to thank Carlos Mafra and Humberto Gomez for helpful remarks and e-mail correspondence during the course of this work. The second named author thanks the support provided by COLCIENCIAS through grant number  FP44842-013-2018 of the Fondo Nacional de Financiamiento para la Ciencia, la Tecnolog\'ia y la Inovaci\'on.

\providecommand{\href}[2]{#2}\begingroup\raggedright\endgroup
\end{document}